\documentclass[review]{elsarticle}
	
\usepackage{amsmath}
\DeclareMathOperator*{\argmin}{argmin}
\newcommand{\norm}[1]{\left\lVert#1\right\rVert}

\usepackage{subcaption}
\usepackage{dingbat}

\usepackage{booktabs}

\usepackage{amsthm}
\theoremstyle{plain}

\theoremstyle{definition}

\usepackage{lineno,hyperref}
\modulolinenumbers[5]
\usepackage{color,soul}

\journal{Journal of \LaTeX\ Templates}

\usepackage{multirow}
\usepackage[table,xcdraw]{xcolor}

\definecolor{newcolor}{HTML}{B2FF66}
\definecolor{oldcolor}{HTML}{F77979}

\begin{document}

\begin{frontmatter}
	
	\title{Recommender Systems for Online and Mobile Social Networks: A survey.}
	
	\author[mymainaddress]{Mattia G. Campana\corref{mycorrespondingauthor}}
	\ead{m.campana@iit.cnr.it}
	\author[mymainaddress]{Franca Delmastro}
	\ead{f.delmastro@iit.cnr.it}
	
	\cortext[mycorrespondingauthor]{Corresponding author}
	\address[mymainaddress]{IIT-CNR, Via G. Moruzzi 1, 56124, Pisa, Italy}
	
	\begin{abstract}
		Recommender Systems (RS) currently represent a fundamental tool in online services, especially with the advent of Online Social Networks (OSN).
		In this case, users generate huge amounts of contents and they can be quickly overloaded by useless information.
		At the same time, social media represent an important source of information to characterize contents and users' interests.
		RS can exploit this information to further personalize suggestions and improve the recommendation process.
		In this paper we present a survey of Recommender Systems designed and implemented for Online and Mobile Social Networks, highlighting how the use of social context information improves the recommendation task, and how standard algorithms must be enhanced and optimized to run in a fully distributed environment, as opportunistic networks.
		We describe advantages and drawbacks of these systems in terms of algorithms, target domains, evaluation metrics and performance evaluations.
		Eventually, we present some open research challenges in this area.
	\end{abstract}
	
	
\end{frontmatter}

\section{Introduction}

Recommender Systems (RS) and Online Social Networks (OSN) have established a strong cooperation in the last few years. They both aim at coping with the huge amount of data produced and shared by users through online platforms, trying to maintain a high user engagement.
This cooperation is built upon the advantages that both systems can achieve: the optimization of the recommendation techniques, by exploiting additional content and user characterization derived from OSN, and the increasing request of OSN services' personalization.
RS targeting the social media domain have been already defined in the literature as Social Recommender Systems (SRS) \cite{ricci2010mobile,Guy:2011:SRS:1963192.1963312}. 
However, this notion should be further expanded, embracing also the recent evolution of social media towards the pervasive and mobile computing environment.
Here, the network can be also implemented by the physical co-location of users and devices, and their ability to exploit direct wireless communications for content sharing and dissemination, without requiring a constant Internet access.
In this direction, the concept of Mobile Social Networks (MSN) has been introduced in \cite{ContiGMP10}, not as a simple extension of OSN services running on mobile devices, but as the opportunity to create real {\it networks of people}, in which users actively participate in the generation and sharing of contents, anywhere and anytime, based on opportunistic networking and device-to-device (D2D) communications.
In this scenario, RS can be optimized by exploiting additional information derived from users' mobile devices (i.e., sensors), which help contextualize the user's preferences, and to rely on a partial and dynamic knowledge of the network and the available data.
This concept contributes to the migration towards a new Internet paradigm, the Internet of People (IoP)~\cite{conti2017internet}, in which users are at the center, by actively contributing to the network evolution both from the communication and data point of views.

In this paper we present a survey of SRS defined both for OSN and MSN, highlighting advantages and drawbacks of standard recommendation techniques applied in these environments, and how those systems evolved over time, in terms of technical solutions, target domains, evaluation metrics and performance evaluations.
In Section \ref{sec:reccomendation_task} we summarize the problem of recommendations and the evaluation metrics generally used in the literature.
Then, in Section \ref{sec:main_approaches}, we present an overview of standard techniques like Collaborative Filtering (CF), Content-based RS, Network-based RS and Context-aware RS, with particular attention to the various information used to characterize the relationship between users and items to further optimize and personalize the recommendations.
This section would also introduce the reader with standard notations and methods applied in SRS, which will be analyzed in the subsequent sections. 
In Sections \ref{sec:osn} and \ref{sec:msn}, we describe the main solutions presented in the literature for OSN and MSN, respectively.
The first area has been widely studied in the last years, proposing solutions that can address different target domains (e.g., to recommend people, locations, POIs, tags or contents) by exploiting heterogeneous context information.
Therefore, we decided to group the proposed solutions by the type of context information used (e.g., social relationships, tags, location) and the recommendation target (e.g., to recommend contents, tags, friends, people).

The research area of RS for MSN is still in its infancy, and few solutions have been presented in the literature, mainly aimed at optimizing content dissemination in opportunistic networks through personalized recommendations.
The main difference between RS for OSN and MSN relies on the knowledge they use to select the recommendations for their final users.
In OSN, RS assume to have access to a complete knowledge of all the available objects on the network (i.e., contents, tags, etc.), residing on a centralized infrastructure.
Instead, in MSN, RS can rely on the local knowledge of each user, represented by the local available objects and those declared by other users in proximity through D2D communications.
In this scenario, each mobile device has a different knowledge of the network, which grows up through its personal mobility and its opportunities to communicate with the others.
In addition, mobile devices have limited computational capabilities and RS must be characterized by efficient response time.
This is due to the limited and unpredictable duration of the opportunistic contact, during which mobile devices can exchange their local knowledge and the recommended objects. 
In order to evaluate and compare the performance of RS for MSN it is necessary to reproduce the realistic behavior of mobile users in a synthetic environment, since a common evaluation framework and appropriate real datasets are not currently available.
In Section~\ref{sec:msn} we present the proposed RS in this area, highlighting the advantages of using context information to improve the recommendation process and demonstrating their efficiency with respect to centralized solutions.
Eventually, in Section~\ref{sec:conclusions} we present some concluding remarks and open research challenges in this area.

\section{The recommendation task}
\label{sec:reccomendation_task}

\begin{figure}[t]
	\centering
	\includegraphics[width=0.98\textwidth]{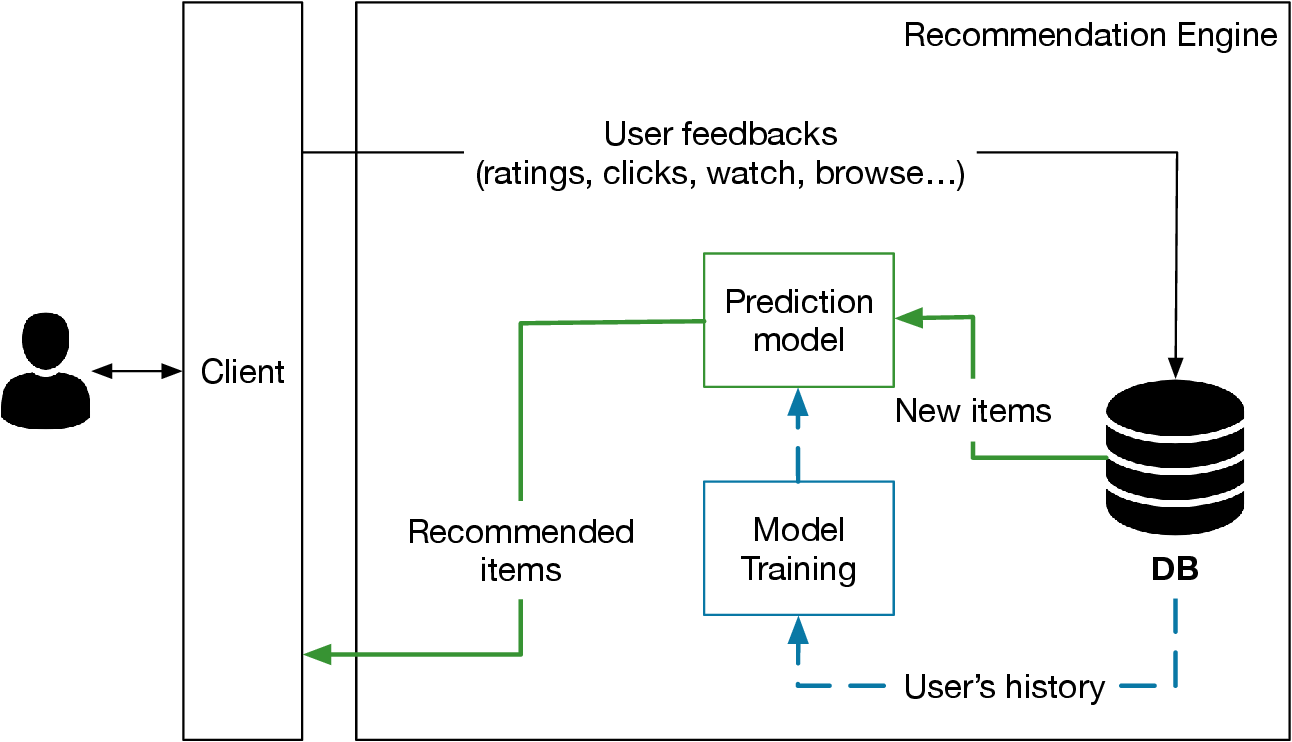}
	\caption{Recommender System architecture}
	\label{fig:recys}
\end{figure}

As a general concept, RS try to identify and foresee users' interests in specific contents, based on their previous experiences.
Figure~\ref{fig:recys} depicts the high-level architecture of a typical RS.
When a user interacts with the system, she provides a set of explicit or implicit feedbacks (e.g., likes, clicks, ratings) about her tastes.
For example, if a user positively rates an article about a new smartphone, she may also be interested in reading news about mobile apps.
Therefore, the basic idea of RS is to exploit this information to infer user interests.
Based on the user's past feedbacks, RS learn a model to predict how much a user can be interested in new items.
Those items are then ranked according to their predicted relevance for the user.
Finally, the higher ranked items will be proactively suggested to the user.

The relationship between users and items is generally represented by a matrix $\mathbf{R_M}$ storing a rating $r_{u,i}$ to identify how much the user $u$ liked the item $i$ in the past. More formally, the matrix $\mathbf{R_M}$ is called \emph{ratings matrix} and it can be defined as follows:

\begin{equation}
	\mathbf{R_M} : U \times I \to R,
	\label{eq:traditional_rec_function}
\end{equation}

where $U = \left\{ u_1, \cdots , u_m \right\}$ represents the set of users, $I = \left\{ i_1, \cdots , i_n \right\}$ is the set of items, and $R = \left\{ r_1, \cdots , r_k \right\}$ denotes the set of possible ratings that users can use to express their grade of interest for a specific item.

The semantic value of the \emph{ratings} highly depends on the application domain. They are often specified as a discrete set of ordered numbers (e.g., a 5-point rating scale is commonly used in e-commerce websites, like Amazon, and video-on-demand services, like Netflix), or a binary value, as in most of OSN (i.e., ``like'' on Facebook or a ``retweet'' on Twitter). In the first case, the ratings are considered as \emph{explicit feedbacks} of the users, while in the latter they are considered as \emph{implicit feedbacks}, derived from the users' activities in OSN. In this paper, we generally refer to ratings scale including both cases.

A missing value in the ratings matrix can have two meanings: (i) the user did not want to express an opinion about a specific item, or (ii) the user does not know that item yet.

As we will discuss in details in the next sections, a great number of recommendation techniques (e.g., Collaborative Filtering) aims to predict missing ratings in $\mathbf{R_M}$ in order to generate lists of new items to recommend to the users.
However, there are also other Recommender Systems (e.g., some network-based RS) that simply provide ranked lists of items without explicitly predicting ratings, using the ratings matrix just to model the users' preferences.

Typically, there are a lot of missing values in $\mathbf{R_M}$. In \cite{shi2014collaborative} the sparsity of the ratings matrix has been calculated to be generally larger than 99\% in commercial systems.
This is due to the \emph{long tail} property~\cite{park2008long}, which is satisfied by different real-world settings.
According to this property, just a small subset of the contents in the system are associated with a high number of rates (i.e., \emph{popular items}) and, consequently, the vast majority of the items are rarely rated by the users.
This characteristic of the ratings matrix has an important implication for RS: the system may not have enough information to generate relevant recommendations for the target user and the presence of popular items may bias the recommendation process, likely providing trivial recommendations.
The main recommendation approaches presented in Section~\ref{sec:main_approaches} address this problem in different ways. However, before detailing them, we briefly describe the most used evaluation metrics.

\subsection{Evaluation metrics}
RS performance can be measured in terms of accuracy in predicting ratings or improvement of the user's experience. This choice depends on the RS target: to increase the profit of the service or to improve the user's satisfaction.

In addition, RS can be evaluated through \emph{online} or  \emph{offline} approaches. 
The \emph{online} approach mainly leverages on the continuous interaction between the users and the system; the user selects an item among those recommended, and this choice is used as input to machine learning algorithms used to adapt the RS behavior to the user's preferences. 
This approach requires that a large number of users is involved in the RS evaluation, and only few research works explored this kind of evaluation until now (e.g., ~\cite{mahmood2007learning, mahmood2009improving, taghipour2007usage}). 
Most RS are currently evaluated using the \emph{offline} approach by exploiting  available datasets including the history of the users actions.
In these cases, the datasets are generally divided into two subsets, namely the \emph{training} ($T_r$) and \emph{test} ($T_e$) sets.
The former represents a set of examples used to fit the parameters of the proposed recommendation model, and the latter is used to test its accuracy.

\subsubsection{Prediction accuracy}
The prediction accuracy measures how much the predicted ratings, derived from the learning phase of the RS, differs from the actual ratings computed on the test set  $T_e$.
More formally, let $r_{jz} \in T_e$ be the explicit rating given by the user $u_j$ for the item $i_z$ in the test set, and $\hat{r}_{jz}$ be the rating predicted by the RS for the user $u_j$ related to the item $i_z$, the accuracy is determined by the error computed for the user-item couple $(u_j, i_z) \in T_e$ as $e_{jz} = \hat{r}_{jz} - r_{jz}$.
Leveraging on the single $e_{jz}$ errors for each pair $<user,item>$, different accuracy metrics have been proposed to compute the overall error over the entire test set $T_e$.
The most used are \emph{Root Mean Squared Error} (\emph{RMSE})~\cite{willmott2005advantages} and  \emph{Mean Absolute Error} (\emph{MAE})~\cite{willmott2005advantages}.
However, they both depend on the specific rating scale used in the reference dataset. Thus, in order to compare the performance of a RS over different datasets, we should use their normalized versions:  \emph{Normalized RMSE} (\emph{NRMSE}) and  \emph{Normalized MAE} (\emph{NMAE}), defined as follows: 

\begin{equation}
	RMSE = \sqrt{\frac{\sum_{(u_j, i_z) \in T_e} e^2_{jz}}{\left| T_e \right|}}; NRMSE = \frac{RMSE}{r_{max} - r_{min}},
\end{equation}

\begin{equation}
	MAE = \frac{\sum_{(u_j, i_z) \in T_e} \left | e_{jz} \right |}{\left| T_e \right|}; NMAE = \frac{MAE}{r_{max} - r_{min}},
\end{equation}

where $r_{max}$ and $r_{min}$ are, respectively, the maximum and minimum ratings in the dataset.
Both RMSE and MAE (and their normalized versions) are meaningful accuracy measures: smaller values, better RS performance.
However, as RMSE sums the squared errors, it may be highly affected by outliers. This means that a few wrong predictions could lead to a worst overall RMSE.


\subsubsection{Ranking accuracy}

Ranking accuracy deals with the different levels of utility of the recommended items with respect to their position in the ranked list proposed to the user.
Ideally, the best ranking in the recommendation list should be assigned to the items considered as the most useful to the user. 
One of the most popular metrics used to evaluate the ranking accuracy is the \emph{Discounted Cumulative Gain} (\emph{DCG}) \cite{Baltrunas:2010:GRR:1864708.1864733} and its normalization, \emph{Normalized Discounted Cumulative Gain} (\emph{NDGC})~\cite{balakrishnan2012collaborative}, which are defined as follows:

\begin{equation}
	DCG = \frac{1}{m} \sum_{u = 1}^{m} \sum_{j \in I_u} \frac{g_{uj}}{log_2(v_j + 1)}; NDCG = \frac{DCG}{IDCG},
\end{equation}

In DCG formula $m$ is the number of users in the test set $T_e$, $I_u$ represents the set of items rated by the user $u$, and $v_j$ is the position of the item $j$ in the recommended list.
The numerator, $g_{uj}$, represents the utility (i.e., the gain) of the item $j$ for the user $u$, and the denominator represents a discount factor with respect to the item position in the ranking list.
The utility directly derives by the ground-truth (i.e., the actual rating $r_{uj}$ provided by the user in the test set).
The Ideal value IDCG is computed with the same DCG formula by using the ground-truth ranking.

Generally, a RS suggests to the users the top-k elements of the recommendation list.
An alternative way to evaluate its accuracy is to consider the trade-off between the length of the list $R_L$ and the number of actual relevant items for the user (i.e., the items positively rated by the user in the test set).

The relevant items contained in $R_L$ are also identified as the \emph{true-positives} (\emph{tp}), while the relevant items not included in $R_L$ are called \emph{false-negative} (\emph{fn}).
In the same way, the proposed items not actually relevant for the target user are called \emph{false-positive} (\emph{fp}), while those not relevant and discarded are called \emph{true-negative} (\emph{tn}).
Therefore, the trade-off between the length of $R_L$ and the number of relevant items can be measured using the \emph{Precision and Recall} measures defined as follows~\cite{herlocker2004evaluating}:

\begin{equation}
	Precision = \frac{tp}{tp + fp}; Recall = \frac{tp}{tp + fn}
\end{equation}

In addition, one way to create a single metric that summarizes both the aforementioned measures is the \emph{F-measure}~\cite{gunawardana2015evaluating}, which is the harmonic mean of equally weighted Precision and Recall:

\begin{equation}
	F = \frac{2 \cdot Precision \cdot Recall}{Precision + Recall}
\end{equation}

\section{Main approaches for Recommender Systems}
\label{sec:main_approaches}

\subsection{Collaborative Filtering}
\label{sec:collaborative_filtering}

Collaborative Filtering (CF) is considered the most popular and widely implemented technique in RS~\cite{ricci2011introduction}.
The underlying assumption of CF is that people with similar preferences will rate same objects with similar ratings~\cite{tang2013social}.
Existing CF solutions can be categorized in two main classes: (i) memory-based and (ii) model-based methods.
Memory-based solutions leverage on similarities in users' behaviors and preferences to make inferences about missing values in the ratings matrix.
Instead, model-based methods exploit the matrix values to learn a model, similarly to a classifier that trains a model from labeled data.
The learned model is then used to predict the relevance of new items for the users.

\subsubsection{Memory-based}

Memory-based algorithms (also known as \emph{Neighborhood-based}) rely on the notion of similarity among users, or items, to predict the possible interest of a user in items that she has not seen (or rated) before.
In the literature, the memory-based CF solutions are typically divided in two main categories: \emph{user-based} and \emph{item-based}.
The former approach is based on the assumption that similar users typically rate the items in a similar way.
Therefore, a user-based CF predicts the rating that a user $u$ might assign to an item by aggregating the ratings that the most similar users to $u$ have previously given to that item.
Formally, the predicted rating of a user $u$ to the item $j$ can be formulated as follows:

\begin{equation}
	\hat{r}_{u,j} = \frac{1}{K}\sum_{k \in N_u} Sim(u,k) \cdot r_{k,j},
	\label{eq:user-based-CF_prediction}
\end{equation}

where $N_u$ is the set of the $K$ users most similar to the target user $u$; $Sim(u,k)$ represents the similarity between the users $u$ and $k$ for a predefined similarity measure, and $r_{k,j}$ represents the rating made by the user $k$ for the item $j$. 

In contrast with user-based CF, item-based CF focuses on the similarities among items.
It is based on the assumption that similar items are rated in a similar way by the same user.
In this case, the items recommended to user $u$ are ranked by aggregating the similarities between each candidate item and the items that $u$ has rated in the past.
It is possible to formulate the prediction rating for the item-based CF as follows:

\begin{equation}
	\hat{r}_{u,j} = \frac{1}{K}\sum_{k \in N_i} Sim(j,k) \cdot r_{u,k},
	\label{eq:item-based-CF_prediction}
\end{equation}

where $N_i$ represents the set of neighbor items of item $j$, and $Sim(j,k)$ is the similarity value (for a predefined similarity measure) between items $j$ and $k$.

The similarity computation between users (or items) represents a crucial step in memory-based approaches, as it can seriously reduce both the accuracy and the performance of RS~\cite{ning2015comprehensive}.
In the literature, a number of similarity measures have been proposed, but the \emph{Pearson Correlation} (PC) similarity seems to provide the most accurate results~\cite{herlocker2002empirical}.
PC selects just the co-rated items and considers the differences in the mean and variance of the ratings made by two users $u$ and $v$.
PC similarity between two users is given by the following formula:

\begin{equation}
	\label{eq:user-pearson}
	PC(u,v) = \frac{\Sigma_{i \in I_{u,v}} (r_{ui} - \bar{r}_u) \cdot (r_{vi} - \bar{r}_v)}{\sqrt{\Sigma_{i \in I_{u,v}} (r_{ui} - \bar{r}_u)^2 \cdot \Sigma_{i \in I_{uv}} (r_{vi} - \bar{r}_v)^2}},
\end{equation}

where $I_{u,v}$ is the set of items rated by both users $u$ and $v$, $r_{ui}$ is the rating made by user $u$ for the item $i$, and $\bar{r}_u$ is the mean value of the ratings made by user $u$.
On the other hand, PC similarity between two items $i$ and $j$ can be calculated comparing the ratings made by users that have rated both items:

\begin{equation}
	\label{eq:item-pearson}
	PC(i,j) = \frac{\Sigma_{u \in U_{i,j}} (r_{ui} - \bar{r}_i) \cdot (r_{uj} - \bar{r}_j)}{\sqrt{\Sigma_{u \in U_{i,j}} (r_{ui} - \bar{r}_i)^2 \cdot \Sigma_{u \in U_{ij}}(r_{uj} - \bar{r}_j)^2}},
\end{equation}

where $U_{i,j}$ is the set of users who rated both items $i$ and $j$, and $\bar{r}_i$ is the average rating value obtained by item $i$.
For a complete comparison of the similarity measures typically used in memory-based CF, we refer the reader to~\cite{ning2015comprehensive}.

The main drawback of memory-based CF is represented by the computational cost: the computation of similarities between all pairs of users, or items, typically requires a quadratic time.
However, they gained popularity due to their very simple implementation, providing an intuitive justification for the computed predictions.
For instance, in item-based CF, the list of similar items can be presented to the user in order to better understand the reason of the recommendation.

\subsubsection{Model-based}
\label{sec:mode-based_CF}
Although memory-based CF approaches are easy to implement and useful in effectively predicting missing ratings, model-based solutions typically show more accurate results~\cite{aggarwal2016recommender}.
The main characteristic of model-based CF is the use of machine learning techniques to learn models that are able to predict the missing ratings.
In the last few years, different models have been proposed for model-based CF, such as Association rule-based~\cite{mobasher2001effective, shyu2005collaborative}, Bayesian networks~\cite{miyahara2000collaborative, su2006collaborative}, Support Vector Machines~\cite{xia2006support}, Neural networks~\cite{salakhutdinov2007restricted} and, more recently, Deep learning methods~\cite{he2017neural}.
However, the Matrix Factorization (MF) models~\cite{koren2009matrix} are considered to be the state-of-the-art in recommendation systems due to their advantages with respect to scalability and accuracy ~\cite{aggarwal2016recommender}.

\begin{figure}
	\centering
	\includegraphics[width=0.9\textwidth]{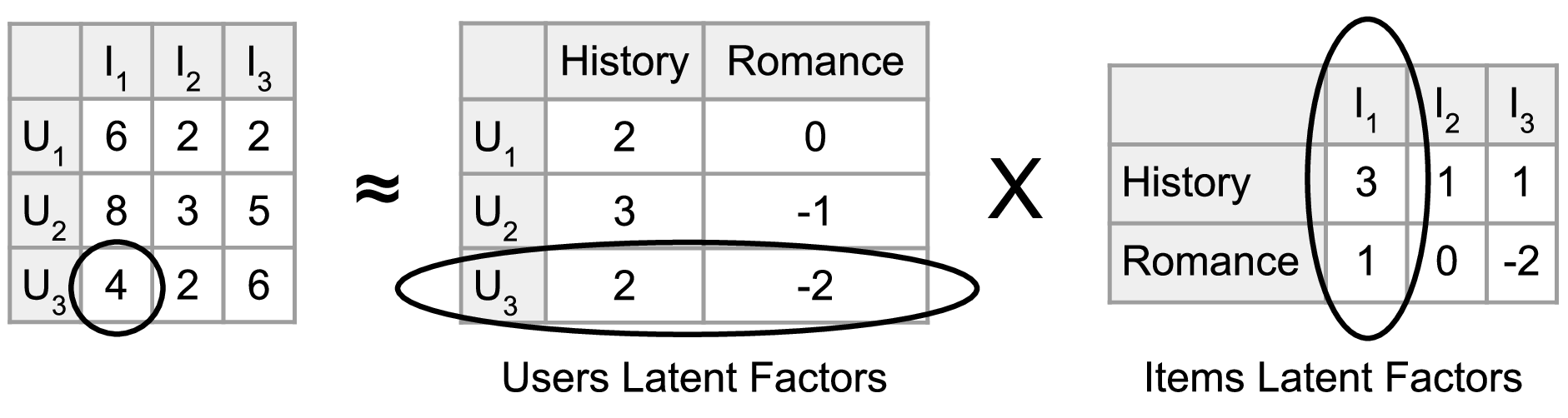}
	\caption{Example of matrix factorization.}
	\label{fig:mf}
\end{figure}

Generally, MF models exploit the typically high correlation between rows (e.g., users) and columns (e.g., items) of an incomplete ratings matrix in order to learn low-rank representations of both users and items.
Moreover, these low-rank matrices (also referred as \emph{latent factors}) can be used to approximate the full rating matrix and then predict missing scores between users and items. 
Specifically, latent factors derived from the correlation patterns in the ratings matrix, which can be used to describe with more details both users and items profiles.
Consider the toy-example depicted in Figure~\ref{fig:mf}.
The ratings matrix contains the preferences of 3 users for 3 different books.
The latent factors may represent the genres of the books.
Therefore, the users' latent factors describe how much each user is interested in a specific genre and, in the same way, the items' latent factors represent how much a book belongs to a specific genre. 

Formally, the matrix factorization method can be expressed as follows:

\begin{equation}
	R \approx U \cdot V^T,
\end{equation}

where $R$ is the full rating matrix, and $U$ and $V$ are two matrices of users and items latent factors, respectively.
The main goal of a MF algorithm is to learn the optimal latent factors ($U^*$ and $V^*$) which better approximate $R$. According to~\cite{shi2014collaborative}, the most common formulation of MF can be formalized as follows:

\begin{equation}
	U^*, V^* = \argmin_{U,V} \left\{ \frac{1}{2} \sum_{i=1}^M \sum_{j=1}^N I_{ij}(r_{ij} - U_i V_j^T)^2 + \frac{\lambda_U}{2} \norm{U}_F^2 \frac{\lambda_V}{2} \norm{V}_F^2 \right\},
	\label{eq:matrix_factorization}
\end{equation}

where $U_i$ is the latent factors of user $i$ and, in the same way, $V_i$ represents the latent factors of item $i$.
$I_{ij}$ denotes an indicator function that is equal to $1$ if rating $r_{ij}$ is not a missing value of the starting rating matrix.
Finally, $\norm{\cdot}_F^2$ represents the (squared) Frobenius norm of a matrix, and $\lambda_U$ and $\lambda_V$ are regularization parameters to set in order to prevent the model's overfitting.
$U$ and $V$ can be seen as unknown variables, which need to be learned in order to minimize the objective function (i.e., Equation~\ref{eq:matrix_factorization}).
This is typically achieved with the well-known Gradient Descent algorithm or one of its variants (e.g., Stochastic Gradient Descent - SGD).

Compared to memory-based approaches, the predictions' accuracy is not the only advantage of model-based CF.
In fact, space requirements for model-based CF are often lower than those required by memory-based algorithms.
This is due to the fact that memory-based approaches should load into memory all the ratings in order to perform the recommendations, while the model learned by model-based CF is typically much smaller than the original ratings matrix.
However, learning a model may require lots of training data and time.
In addition, if the system is highly dynamic (e.g., new users or items are frequently added), it could be necessary to train a new model several times since the current one can easily become obsolete, thus affecting its accuracy.
This could be an issue for some specific application domains. For example, in a Mobile Social Network, each node (e.g., mobile device) has a limited knowledge of the items (and users) available in the network, and it could be difficult to train an accurate model.

\subsection{Content-based Recommender Systems}

CF approaches can be considered inherently ``social'' since they exploit the correlations in the ratings patterns among users of the same system.
However, CF methods typically ignore the item attributes to compute the predictions in their recommendation process, while they can be useful to improve the recommendations' accuracy.

\emph{Content-based Recommender Systems}~\cite{pazzani2007content, lops2011content} are specifically designed to recommend items similar to the ones that the user has preferred in the past. In CF the similarity between two items (or two users) is calculated in terms of correlation or similarity among ratings provided by \emph{other} users. Instead, content-based approaches consider only the ratings provided by the target user and the features of the rated items~\cite{aggarwal2016recommender}.
Content-based RS are typically implemented using traditional classification and clustering algorithms, such as Support Vector Machines (SVM) or Nearest Neighbors methods (NN)~\cite{lops2011content}.
In SVM the features of the items and the ratings made by the target user are combined together to form a dataset of $\left< features, rating \right>$ instances.
Based on this dataset, a specific classifier (or regressor) for the target user is trained in order to predict the ratings of new items, never seen before by the user.
In NN, in order to classify a new, unrated item, the algorithm compares it with all the stored items, derived from the training data, by using a similarity function (typically the Euclidean distance or the Cosine similarity) .

According to Aggarwal~\cite{aggarwal2016recommender} and Tang et al.~\cite{tang2013social}, content-based RS have the following drawbacks:
\begin{enumerate}
	\item \emph{Items' features}. The accuracy of RS relies on the set of features that describe the items.
	The identification of the most relevant features is not trivial, and highly depends on the specific application.
			
	\item \emph{Over-specialization}. Since the content-based methods rely just on the characteristics of the items already rated by the target user, it typically suffers of low diversity and novelty of the recommended items.
	Thus, a content-based RS typically tends to provide obvious recommendations, which could bother the user in the long term.
			
	\item \emph{Training set size}. In order to allow a content-based RS to learn the user's preferences, the user has to rate a sufficient number of items~\cite{tang2013social}.
	Otherwise, RS does not has enough information to learn an accurate model and fails to recommend items for a user with few or no ratings.
\end{enumerate}

Despite the aforementioned drawbacks, content-based systems are useful to alleviate some critical problems of CF.
For example, when a new item is added to the ratings matrix, CF methods are not able to perform recommendations about this item because the system has not yet collected sufficient ratings about it.
In the literature, this problem is called \emph{cold-start}, and a content-based approach can well complement a CF system because of its ability to exploit the characteristics of the items in the recommendation process~\cite{aggarwal2016recommender, miranda1999combining}.
Mixing two or more approaches is often referred as \emph{Hybrid RS}~\cite{adomavicius2005toward} in which the goal is to combine the strengths of different methods to create a more robust RS.

\subsection{Graph-based}
\label{sec:network-based}

Social Networks can be naturally modeled as graphs, in which nodes represent users and items, and edges model the different relationships among user-user or user-item pairs (e.g., friendships, follow, likes, share). These relationships can also be used in RS to identify similar users and/or items. In fact, several graph-based RS have been recently proposed in the literature~\cite{yu2016network}.
Their first objective is to perform a ranking of the nodes to identify the most interesting items for a user or a neighborhood (e.g., the most similar users/items) to be used in a traditional CF method.

PageRank~\cite{page1999pagerank} is clearly the most famous method to produce a ranking of the nodes in a graph.
It is based on the idea that a node is important if it is linked to other important nodes.
This simple and general idea can be applied to many different situations, including ranking of websites in a search engine, or finding relevant items (or people) in a RS.
PageRank exploits a random-walk model.
Specifically, it performs an exploration of the given graph by visiting nodes following randomly selected links among them.
The long-term frequency of visits to a node represents its final ranking value, which is clearly influenced by the number of incoming links of the node.
It is also referred to as the \emph{steady-state probability}.
However, directed graphs can have some nodes (or group of them) without out-going links; this part of the network would act as a trap for the random-walk process.
In order to overcome this problem, PageRank uses the so called \emph{restart} mechanism.
According to this strategy, at each transition, the random walk may either jump to an arbitrary node of the network with a probability $\alpha$, or follow one of the out-going links connected to the current node with probability $(1-\alpha)$.
Formally, the steady-state probability at a node $i$ is defined as follows:

\begin{equation}
	\pi(i) = \frac{\alpha}{n} + (1 - \alpha) \cdot \sum_{j \in In(i)} \pi(j) \cdot p_{ji},
\end{equation}

where $n$ is the total number of network's nodes, $In(i)$ is the set of nodes that have out-going links directed to $i$, and $p_{ji}$ is the transition probability from node $j$ to node $i$. 

Although PageRank is very effective in finding popular nodes in the graph, it does not take into account the user's preferences, which means that it is not able to provide personalized recommendations.
To this aim, Haveliwala~\cite{haveliwala2002topic} proposes \emph{personalized PageRank}, a biased version of PageRank, specifically designed to find popular nodes which are also similar to specific nodes in the graph. 
In this case, the key idea is to use the restart mechanism to bias the random walk process towards specific nodes that represent the user's interests.
If we consider a specific OSN represented as a graph, nodes represent the items available in a social media, and the links represent the relations between users and items. The restart nodes represent the items consumed by the target user in the past.
Then, exploiting the biased random-walk, personalized PageRank provides a way to calculate similarity scores of nodes based on their structural similarity to the restart nodes.
Finally, the items with higher similarity values can be recommended to the target user.

Even if (personalized) PageRank and other similar random walk-based algorithms (e.g., \emph{FolkRank}~\cite{hotho2006folkrank}, \emph{SocialRank}~\cite{tsai2014ranking}, and \emph{ItemRank}~\cite{gori2007itemrank}) can be directly employed to build Social Recommender Systems (SRS)~\cite{tsai2014ranking,lakiotaki2011multicriteria}, they are mostly used to select user/item's neighbors in memory-based CF solutions~\cite{fouss2007random, yildirim2008random}.
In fact, rather than using standard similarity measures (e.g., Pearson Correlation Coefficient on user's ratings), it is possible to use such structural measures to identify the most relevant users (or items) for the recommendation purpose.
Since they only leverage on the structural transitivity of the network's edges, instead of the users' ratings, they typically result very effective to alleviate the problems related to the sparsity of the ratings matrix.
However, random walk-based approaches typically require multiple iterations before the transition probabilities converge to a steady state.

On the contrary, other solutions require only few steps to generate effective recommendations~\cite{zhou2010solving}.
\emph{Diffusion-based} methods~\cite{yu2016network} represent the most used approach for recommendation in graph-based RS. They have been initially proposed by Zhou et al.~\cite{zhou2007bipartite,zhou2010solving} with their algorithms \emph{ProbS} (Probability Spreading)~\cite{zhou2007bipartite} and \emph{HeatS} (Heat Spreading)~\cite{zhou2010solving}.
These algorithms exploit a bipartite user-item graph where users are directly connected to the items collected in the past.

\begin{figure}[t]
	\centering
	\includegraphics[width=0.9\textwidth]{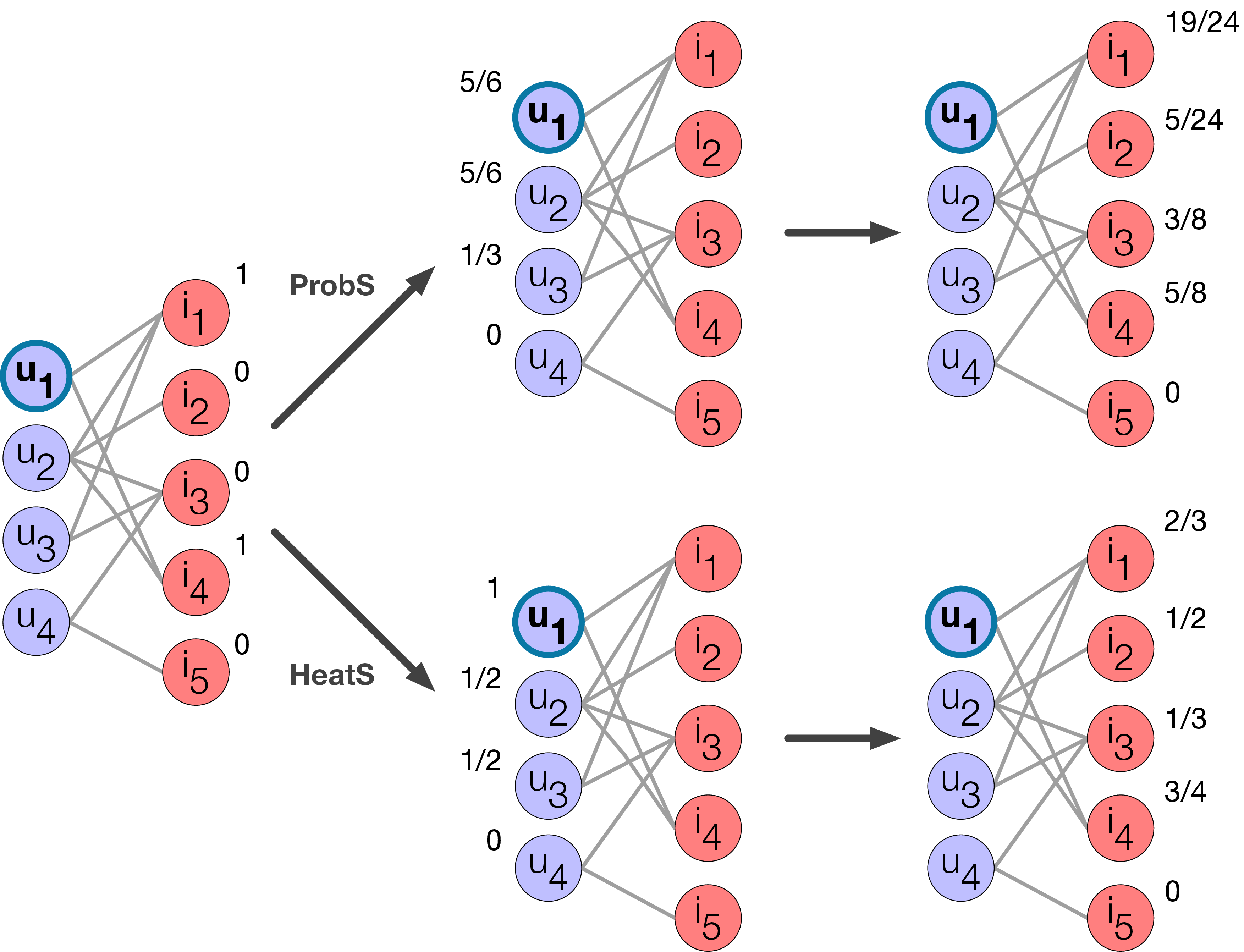}
	\caption{Execution of ProbS and HeatS on a bipartite user-item graph.}
	\label{fig:probs_heats}
\end{figure}

In ProbS, a resource is initially assigned to each item connected to the target user, initialised to a unitary value.
This resource is then spread uniformly from the collected items to the users connected with them (\emph{first diffusion step}) and then, in a second step back, to the items connected with those users (\emph{second diffusion step}).
The final value of this resource associated with each item is then interpreted as the recommendation score of that item for the target user.
The higher the score obtained by an item, the greater could be the interest in it by the target user.

HeatS~\cite{zhou2010solving} is similar to ProbS, but it is based on opposite rules: each time the resource (or a portion of it) is redistributed, it is divided by the number of edges connected to the node towards which it is heading to.
Figure~\ref{fig:probs_heats} depicts the diffusion steps of the two algorithms, highlighting the differences in the two recommendations.

Unfortunately,  both of them are actually biased by the presence of extremely popular or non-popular items (i.e., nodes with very high or low degree, respectively), not taking into account the characteristics of the users' interests.
To overcome these limitations, a ProbS+HeatS hybrid approach, simply called \emph{Hybrid}, has been proposed in~\cite{zhou2010solving}.
Hybrid calculates a linear combination of the results of ProbS and HeatS with a parameter $\lambda$ governing the relative importance of one of the two original algorithms.
By tuning $\lambda$ appropriately, Hybrid is able to obtain simultaneous gains in both accuracy and diversity of recommendations with respect to the previous solutions.

The diffusion-based approach has been generalized in multiple ways. Some authors proposed to compensate the HeatS's preference for unpopular items modifying its spreading process~\cite{liu2011informationbiased} and allocating higher weight to links connecting nodes with high degree~\cite{liu2011information}.
On the other hand, in order to compensate the ProbS's bias towards very popular items, L{\"u} et al.~\cite{lu2011information} proposed to highlight less popular items in the second diffusion step by making their score inversely proportional to the degree of the nodes.

Furthermore, Zhou et al. in ~\cite{zhou2008effect} investigated the effect of biasing the initial resource allocation in ProbS.
Results indicate that decreasing the initial resource allocated on popular objects can further improve the algorithmic accuracy. Similarly, Liu and Zhou used heterogeneous initial resources in Hybrid, further improving both the accuracy and diversity of the recommendations~\cite{liu2012heterogeneity}. However, in addition to the ProbS-HeatS hybridization parameter $\lambda$, they introduce a second parameter to tune, which governs the heterogeneity of the initial resources.

For a complete comparison and evaluation of the diffusion-based methods, we refer the reader to~\cite{yu2016network}.


\subsection{Context-Aware Recommender Systems}
\label{sec:cars}

The RS we have presented focus on recommending the most relevant items to individual users taking into account just two entities: users and items they liked in the past.
However, they do not consider additional information that can be useful to improve the recommendations' quality and personalization~\cite{shi2014collaborative}.
To this aim, \emph{Context-Aware Recommender Systems} (\emph{CARS})~\cite{adomavicius2011context} have been recently proposed in the literature.
CARS methods perform their recommendations considering also the information characterizing the specific situation in which the target user is involved.
This additional information is referred to as the \emph{context}.
The notion of context has been studied in multiple disciplines and several heterogeneous definitions exist~\cite{adomavicius2011context}.
In CARS context is defined as the additional information relevant to improve recommendations, such as the time of the day, the locations of the target user, and her social relationships.

\begin{figure}[t]
	\centering
	\includegraphics[width=0.6\textwidth]{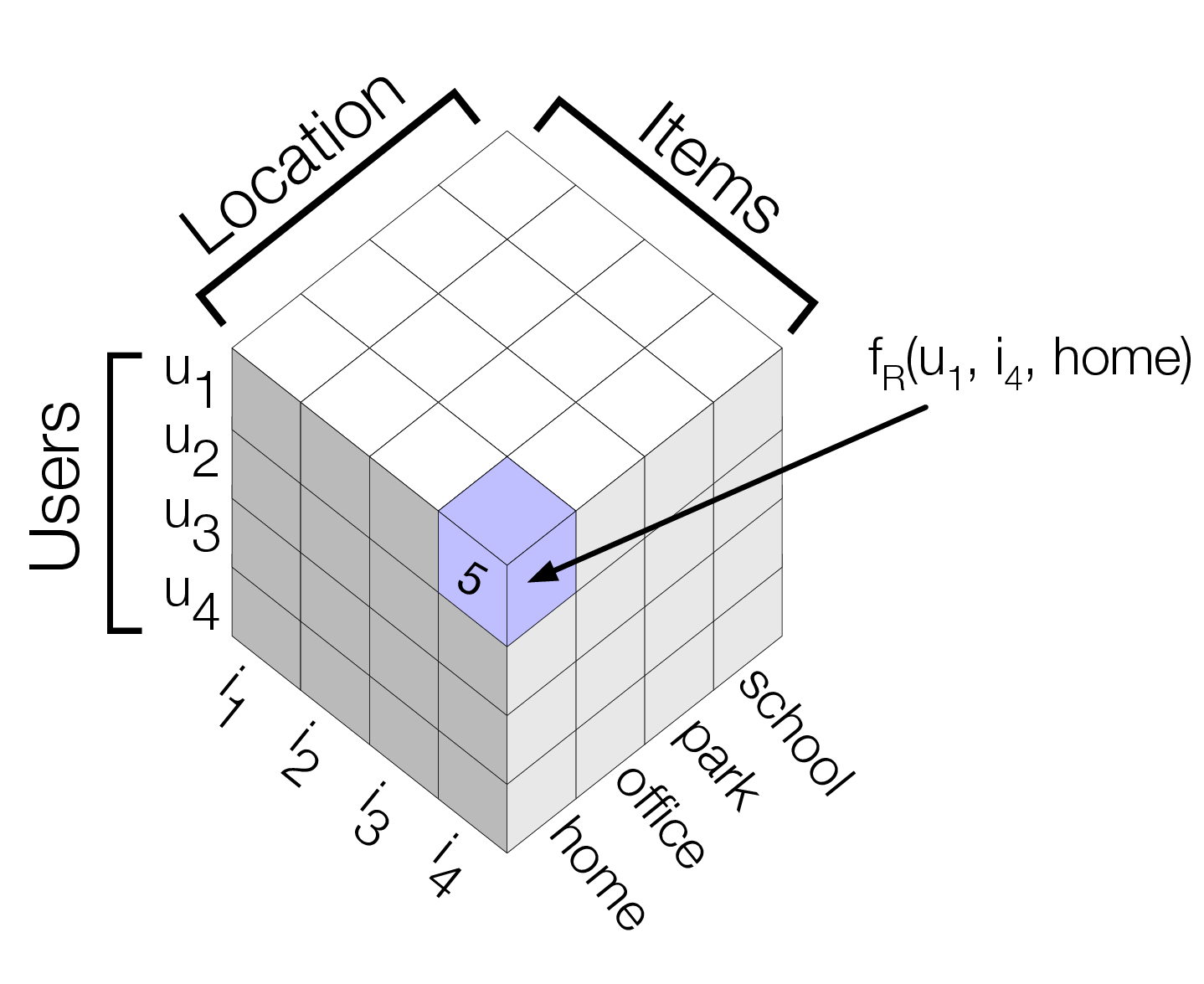}
	\caption{Example of a multidimensional cube for the $User \times Item \times Location$ recommendation space.}
	\label{fig:tensor}
\end{figure}

As we described in Section~\ref{sec:reccomendation_task}, the objective of a traditional RS is to approximate the rating's function $f_R: U \times I \rightarrow R$ in order to predict the possible interest of a user for a specific item.
CARS generalize this approach by using a \emph{multidimensional approach}~\cite{adomavicius2005incorporating} in which the rating function can be seen as a mapping from a \emph{n}-dimensional tensor (i.e., a multidimensional matrix) to the set of ratings $R$:

\begin{equation}
	f_R : D_1 \times \ldots \times D_n \to R,
\end{equation}

where two dimensions, typically, are the sets of users and items (i.e., $U$ and $I$), and the other dimensions represent the context variables.
Grafically, CARS's rating function can be represented as a multidimensional cube.
Figure~\ref{fig:tensor} shows an example with a 3-dimensional cube that stores the ratings for the recommendation space $User \times Item \times Location$, where
$f_R(u_1,i_4,home) = 5$ means that the user $u_1$ rated with a score of 5 the item $i_4$ when she was at home.

The paradigms used to incorporate context information in RS are divided in three main categories~\cite{adomavicius2015context}: i) \emph{Contextual pre-filtering}, ii) \emph{Contextual post-filtering}, iii) \emph{Contextual modeling}.
In the following we describe in details each approach.

\begin{figure}[t]
	\centering
	\includegraphics[width=0.9\textwidth]{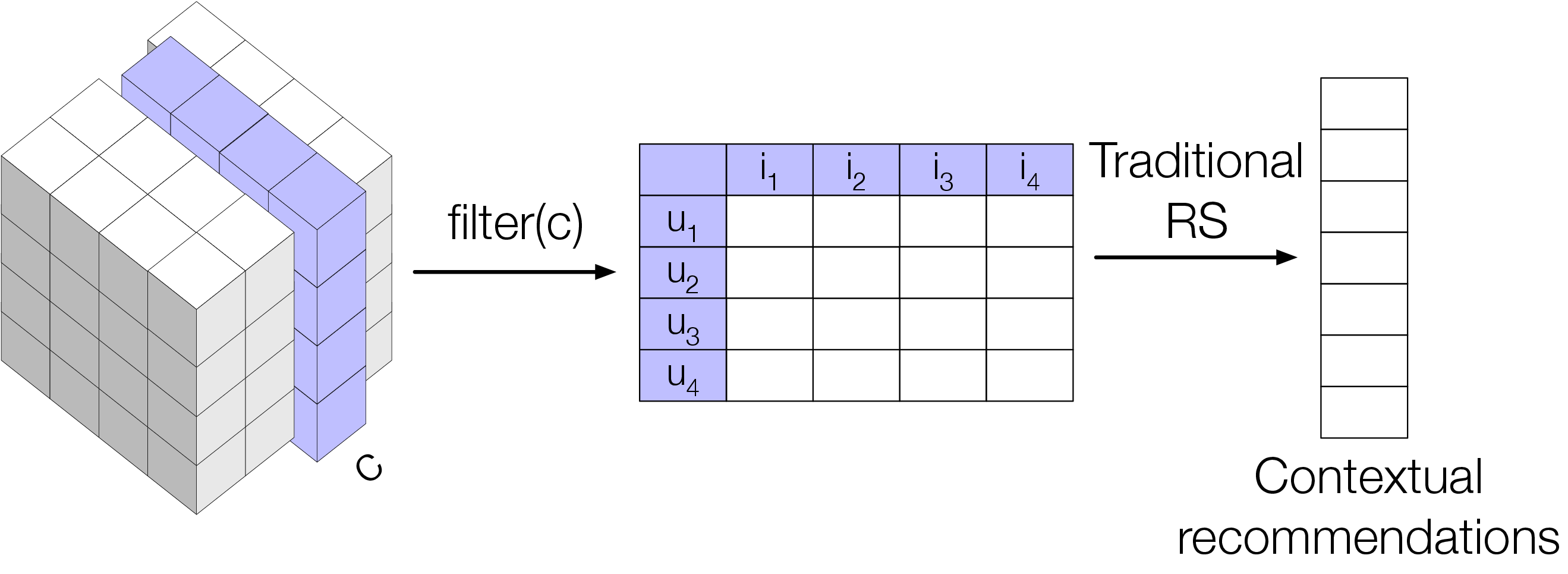}
	\caption{Process of a pre-filtering method to generate contextual recommendations.}
	\label{fig:pre_filtering}
\end{figure}

\paragraph{Contextual pre-filtering}

In the contextual pre-filtering approaches the idea is to reduce the multidimensional cube to a 2-dimensional matrix in order to apply traditional algorithms.
As depicted in Figure~\ref{fig:pre_filtering}, the current context $c$ of the target user is used as a filter to ``slice'' the rating cube and extract the relevant user-item matrix, which can be used with traditional CF algorithms.

However, using the exact context $c$ sometimes can be a problem.
If there are just few ratings associated with $c$, this can lead to a low accuracy in the recommendation because the system has not enough data points about the past preferences of the target user.
Therefore, the notion of \emph{generalized pre-filtering}~\cite{adomavicius2005incorporating} has been introduced in the literature to cope with this problem.
Instead of using the specific context $c$, this approach suggests to use a broader context $c'$, which includes $c$ and other similar context variables.
Let's consider the context of watching a movie on Saturday (i.e., $c = \text{``Saturday''}$).
In order to base the recommendation on a wider set of ratings, the generalized pre-filtering approach may use the more general context $c' = \text{``Weekend''}$, which includes the two context variables $c = \text{``Saturday''}$ and $c_1 = \text{``Sunday''}$.
Then, the system aggregates the selected ratings using some aggregation function (e.g., the average) to reduce the recommendation space to the 2-dimensional user-item matrix and, finally, perform the recommendation.

It is clear that the context generalization becomes a crucial aspect of the recommendation's process.
One possible strategy is to use a process similar to the cross-validation: in a learning phase, the predictive performance of the system is empirically evaluated using different contextual filters.
Then, the filter that allows the best performance is automatically selected~\cite{adomavicius2015context}.
In addition, multiple context generalization can be used for the same specific context.
For this reason, in the literature several works propose to use a combination of different filters and prove that this solution can provide a significant performance improvement compared to the use of a single filter~\cite{adomavicius2015context, burke2002hybrid}.
In this case, multiple pre-filters are used to generate several different 2-dimensional ratings matrices.
Eventually, those user-item matrices are merged in one single matrix in order to perform the recommendation process.
Codina et al. recently proposed in~\cite{codina2013exploiting} the use of semantic similarities between contextual situations in order to select the most relevant context variables for a specific situation.
They showed that their semantic pre-filtering model is able to obtain better results than others context-aware approaches.

\begin{figure}[t]
	\centering
	\includegraphics[width=0.95\textwidth]{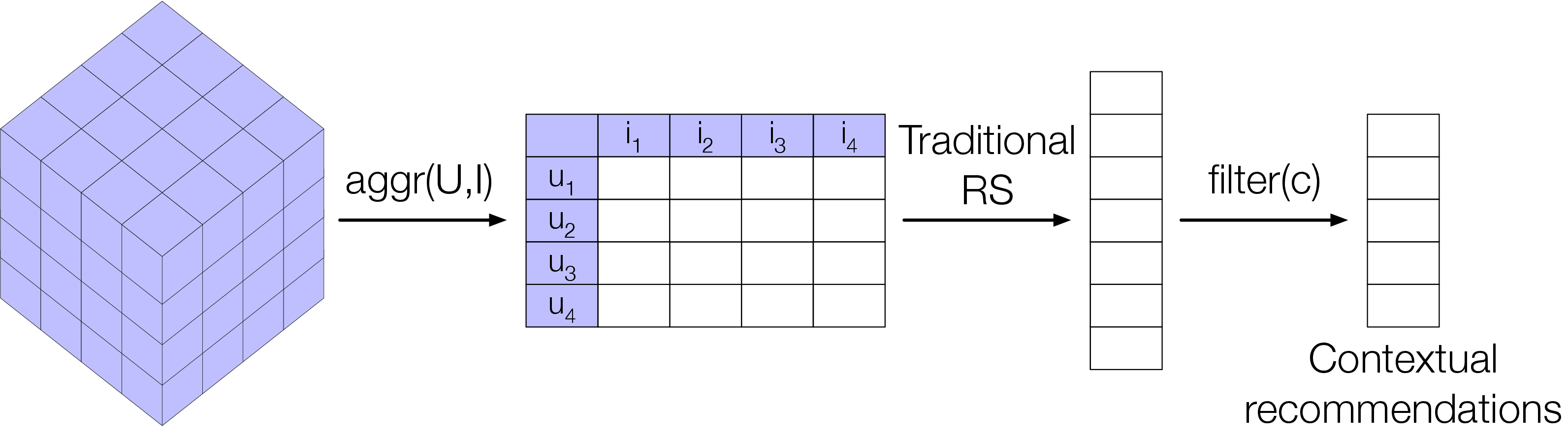}
	\caption{Process of a post-filtering method to generate contextual recommendations.}
	\label{fig:post_filtering}
\end{figure}

\paragraph{Contextual post-filtering}

In this case the filtering step is applied to the output obtained after applying traditional RS to the data set. Therefore, the recommendation process of a post-filtering method can be summarized in two steps:

\begin{enumerate}
	\item In a first stage, the context information are completely ignored, and a list of recommended items are generated using traditional RS on an aggregated 2-dimensional user-item matrix.
			
	\item Given a specific context $c$, recommendations not relevant for $c$ are filtered out, and the ranking of recommendations in the list is adjusted according to $c$.
\end{enumerate}

Figure~\ref{fig:post_filtering} depicts the recommendation process of the post-filtering approach. Here, $aggr(U,I)$ refers to the aggregation function (e.g., the average) used to create the 2-dimensional user-item matrix, and $filter(c)$ represents the filtering (or adjustment) step of the recommendation's list, based on the current context $c$ of the target user.

The selection of the best contextual filter represents a crucial point in the overall recommendation process.
All the techniques related to the context generalization, previously discussed, can be applied also in post-filtering methods.
The main difference between the two approaches lies in the data set on which the contextual filter is applied: the ratings contained in the multidimensional tensor (i.e.,the input) for the pre-filtering approach, and the recommendation list (i.e., the output) generated by a traditional RS for post-filtering methods.

Both approaches share the same advantages: their implementation is relatively simple, and both of them allow the use of the traditional recommendation techniques described in the previous sections.
In addition, comparing their performance, none of them seems to clearly outperform the other. Panniello et al. in~\cite{panniello2009experimental} evaluated different solutions based on both approaches by using several real world datasets. The authors did not find any solution that dominates over the others for all the datasets, indicating that the choose of the best approach really depends on the specific application.

\paragraph{Contextual modeling}

\begin{figure}[t]
	\centering
	\includegraphics[width=0.95\textwidth]{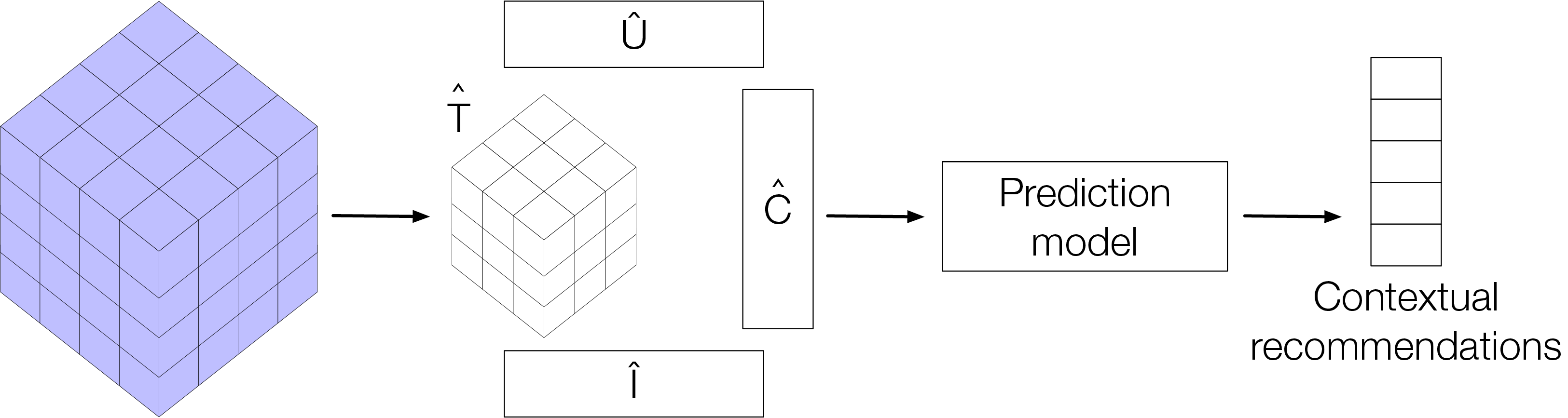}
	\caption{Tensor Factorization.}
	\label{fig:tensor_factorization}
\end{figure}

Contextual modeling approaches consider context information directly in the recommendation function in order to predict a user rating for a specific item~\cite{adomavicius2015context}.
Differently from pre- and post-filtering techniques, contextual modeling methods use predictive models or heuristics to create multidimensional recommendation functions.

The simplest solution adapts neighborhood-based CF methods to perform context sensitive recommendations~\cite{adomavicius2005incorporating}.
In this case, the traditional user-user (or item-item) similarity metric is replaced by an \emph{n}-dimensional distance metric, which includes the contextual information.
For instance, consider two points in a 3-dimensional user-item-context cube, respectively $A = (u,i,c)$ and $B = (u',i',c')$.
The distance between them, $Dist(A,B)$, can be calculated as the sum of the weighted distance between each dimension or, alternatively, using the Euclidean metric~\cite{adomavicius2005incorporating}.
Therefore, by extending the traditional neighborhood rating prediction (i.e., Equation~\ref{eq:user-based-CF_prediction}), the specific rating for user $u$, item $i$ in the context $c$, $r_{u,i,c}$, can be expressed as follows:

\begin{equation}
	r_{u,i,c} = k \sum_{(u',i',c') \neq (u,i,c)} \frac{1}{Dist((u',i',c'),(u,i,c))} \cdot r_{u',i',c'},
\end{equation}

where $k$ is a normalization factor, and the inverse of the multidimensional distance between points has been used for weighting the rating $r_{u',i',c'}$ (i.e., the more the two points are close, and the higher is the weight).

More complex approaches extend model-based solutions.
In particular, tensor factorization methods~\cite{karatzoglou2010multiverse, rendle2010factorization} can be considered as a generalization of matrix factorization for latent factor models (see Section~\ref{sec:mode-based_CF}).
As depicted in Figure~\ref{fig:tensor_factorization}, a 3-dimensional tensor factorization aims at learning three latent factors matrices $\hat{U}$, $\hat{I}$, $\hat{C}$ (i.e., respectively, the user, item, and context latent factors) and a core tensor $\hat{T}$ to build a prediction model.
However, the computational complexity of this kind of models increases exponentially with the number of the tensor's dimensions~\cite{rendle2011fast}.
To solve this issue, some simplified forms of these factorization methods have been recently proposed. For instance, the \emph{Pairwise Interaction Tensor Factorization} (\emph{PITF})~\cite{rendle2011fast} and \emph{Factorization Machines}~\cite{rendle2010factorization, rendle2012factorization} are able to predict missing ratings exploiting the pairwise interactions between the different dimensions of the tensor, maintaining at the same time a low computational complexity.

\begin{figure}[t]
	\centering
	\includegraphics[width=0.98\textwidth]{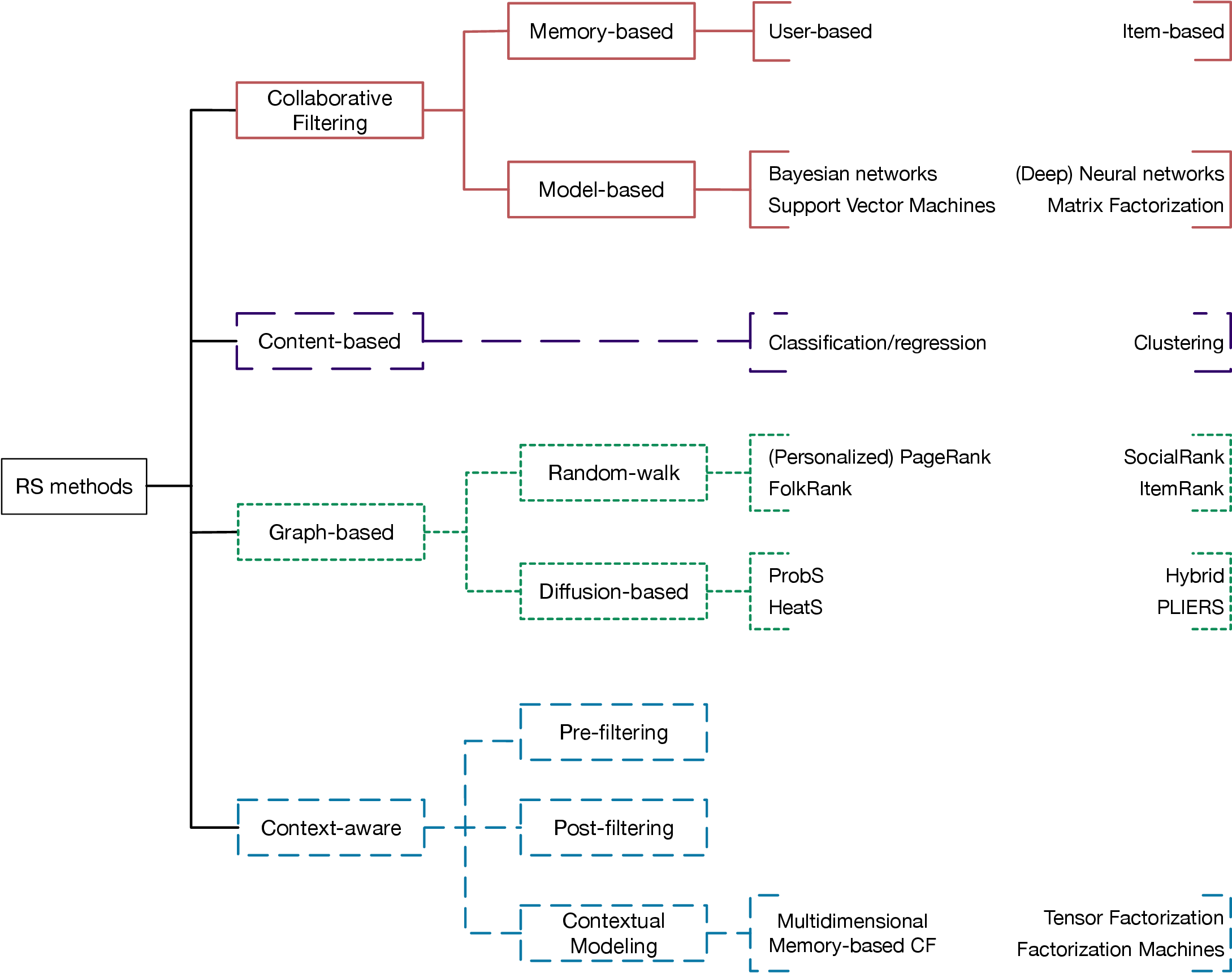}
	\caption{Classification of Recommender Systems methods.}
	\label{fig:RS_classification}
\end{figure}

To conclude the description of main RS methods, we present in Figure \ref{fig:RS_classification} a summary of their classification and the references of the main representative solutions for each method. 

\begin{figure}[t]
	\centering
	\includegraphics[width=0.98\textwidth]{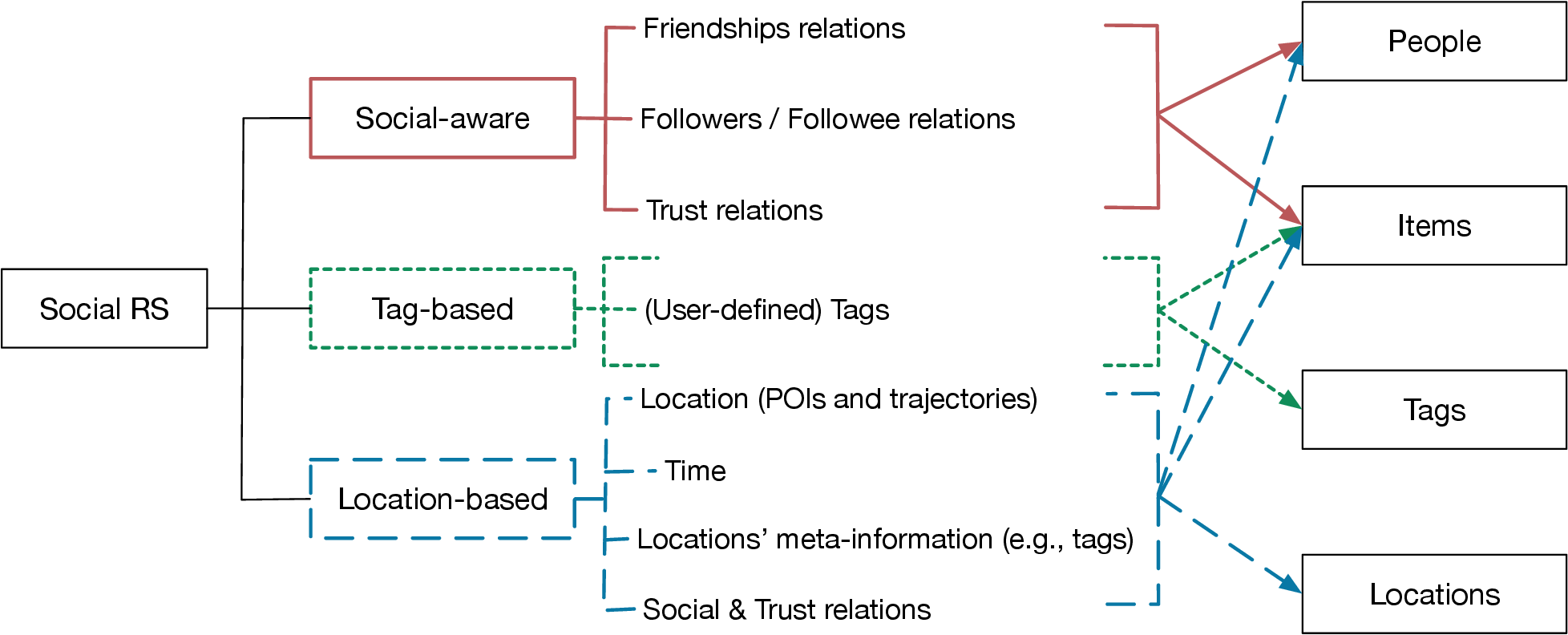}
	\caption{Classification of Social Recommender Systems.}
	\label{fig:srs_classification}
\end{figure}

\section{Recommender Systems for Online Social Networks}
\label{sec:osn}
With the advent of OSN, RS have been further enhanced by exploiting additional information characterizing both users and items, becoming thus SRS~\cite{tang2013social, guy2015social}. Specifically, OSN can provide  information able to identify virtual and physical social relationships among users, common interests and habits, in addition to personal perferences. RS can exploit this information depending on the application domain in which they are implemented. Specifically, SRS widened their recommendation targets, from generic item recommendations up to people/friend recommendations, locations, tags and others. In this section, we provide a description of SRS grouped by the type of recommendations they provide, and the context information used to optimize their process, as shown in Figure \ref{fig:srs_classification}. Specifically, we identify three types of recommendations: (i) \emph{social-aware recommendations}, in case SRS recommend people or friends, and include the notion of trust and social relationships in generic items recommendation; (ii)\emph{tag-based recommendations}, when SRS exploit a specific characterization of items based on \emph{tags} to recommend other items, tags and/or people, e.g., if they share similar interests derived from tags; (iii) \emph{location-aware recommendations}, when SRS exploit location-related information to recommend items, point-of-interest (POIs), trajectories, and people.

\subsection{Social-aware recommendations}
\label{sec:social}

OSN provide several information that characterizes the social context of a user and her generated items. OSN analysis mainly refers to the topological structure of the social network, in terms of (virtual) social relationships among users~\cite{DUNBAR201539}\cite{valerio}. In addition, OSN can reflect the presence of the homophily principle~\cite{mcpherson2001birds} also in the cyber-world, as the tendency of individuals to associate and connect with similar others, as demonstrated by recent studies~\cite{valerio}. 

\begin{figure}[t]
\begin{subfigure}{.5\textwidth}
\centering
\includegraphics[width=.85\linewidth]{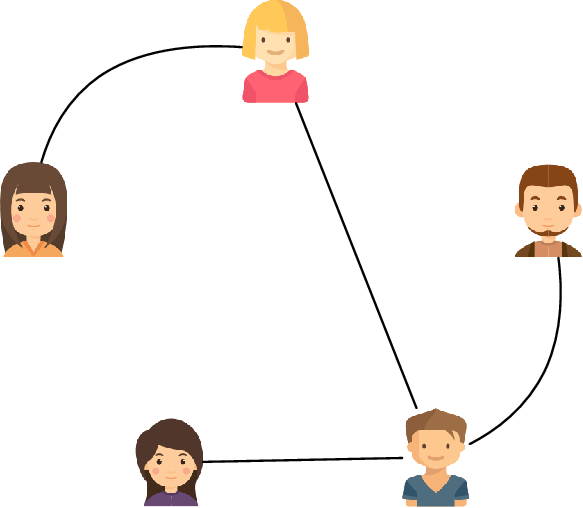}
\caption{}
\label{fig:social-net}
\end{subfigure}%
\begin{subfigure}{.5\textwidth}
\centering
\includegraphics[width=.85\linewidth]{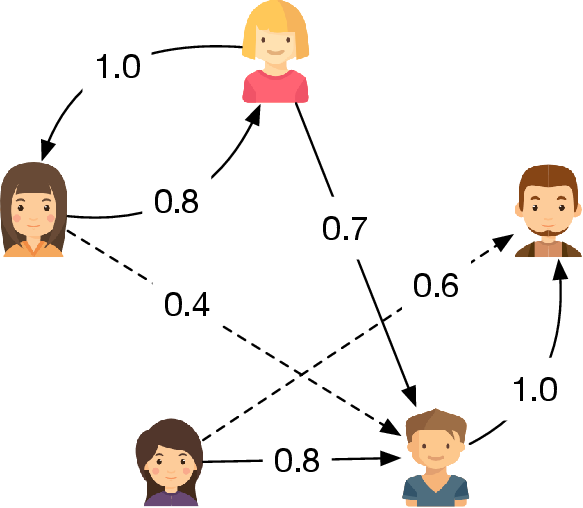}
\caption{}
\label{fig:trust-net}
\end{subfigure}
\caption{Graphical representation of (a) a basic friendship network, and (b) a trust-network with inferred trust links (dashed lines).}
\label{fig:social-trust-nets}
\end{figure}

The social structure of a OSN can be represented as a graph $G = (U, S)$, where $U$ is the set of users and $S$ is the set of social links among them, which can be: i) undirected, if they model friendship relations, such as in Facebook~\footnote{www.facebook.com}, or ii) directed, if they model trust or follow relations, such as in Epinions~\footnote{www.epinions.com} or Twitter~\footnote{twitter.com}.
Social link in OSN represent the explicit declaration of a user to be virtually in contact with others, but they do not reveal anything about the effective nature of the relationships or its strength and value for the user.
However, analysing OSN contents and users' virtual interactions (e.g., explicit users' feedbacks, ratings~\cite{o2005trust}, etc.) we can derive additional information that can further characterize those links, like for example the notion of \emph{trust} among users, which can be represented through links' weights (Figure~\ref{fig:trust-net}). 

SRS exploit this structure to provide two types of social-aware recommendations: (i) friends or other people, and (ii) OSN items. In the first case, SRS suggests to the users new social relationships based on the prediction of common interests, while in the second case it exploits social/trust relationships in order to suggest interesting items to the users.
In the following we review the main solutions proposed in the literature, highlighting different approaches used, and how they take into account different aspects of the social context.

\subsubsection{Friends and People Recommendations}
This recommendation task differs from the generic items recommendation since it must take into account specific aspects of the social network of each user, like trust, reputation, privacy and personal attraction~\cite{guy2015social, terveen2005social}. From a technical point of view, this task is associated with the problem of link prediction in the social network, which tries to infer new possible relationships or interactions between pairs of entities, based on their properties and the existing links~\cite{liben2007link}.

In the literature, several techniques have been been proposed for link prediction~\cite{martinez2016survey}. The most commonly used can be divided into two main categories: unsupervised and supervised methods/heuristics. The first category includes all those techniques that exploit the structure of the network to calculate similarity values among the users (nodes of the graph), and then recommend the most similar ones.
For instance, Jaccard coefficient~\cite{martinez2016survey} or the Adamic-Adar measure~\cite{martinez2016survey} can be used to calculate the similarity between two nodes based on their common neighbors. Alternatively, random walk-based approaches such as Katz measure~\cite{martinez2016survey}, or PageRank~\cite{page1999pagerank} and SimRank~\cite{jeh2002simrank} algorithms, can represent a different way to recommend new links based on the connectivity between pairs of nodes.

One of the first unsupervised RS for people recommendation has been proposed by Guy et al. for an IBM enterprise social network called SONAR \cite{guy2008harvesting, guy2008public, guy2009you}.
It defines a ``relationship score'' based on a combination of various social context information derived from the enterprise intranet applications (e.g., organisational charts, papers database, patents database, a social networking site, a friending system and others).
For each application, they define a user-user graph in which the links' weight is computed with a specific metric depending on the application (e.g., manager-employee relationship, paper/patent co-authorship, users belonging to the same group in the friend network). Then,  a linear combination of the weights of each graph is computed for each pair of users,  and it is used as general relationship score. 
The authors compared SONAR relationship score with that provided by other approaches: (i) \emph{Friend-of-friends (FOF)} metric, applied to the internal friending system of the company; (ii) a \emph{Content-Matching} approach, defining the users' similarity based on a set of keywords extracted only from the contents generated on the social network (i.e., posts, picture descriptions and tags); (iii) \emph{Content-plus-Link}, by doubling the similarity value derived from the Content-Matching in case the pair of users are also explicitly connected in the social network application.
They evaluated SONAR with real experiments involving up to 3000 users in different testing scenarios and demonstrated the effectiveness of the proposed solution with respect to the other approaches.




Other solutions have been proposed by exploiting the concept of \emph{ego-network}, the representation of a social network centered on the individual (i.e., ego) and her social links (i.e., alters) divided in a series of ``circles''. The inner-most circle includes alters with a very strong relationship with the ego. Each subsequent circle (in hierarchy) includes all the relationships of the previous circles along with an additional set of social links with a weaker level of intimacy. The last set, included in the outermost circle only, contains simple acquaintances, with a relatively weak relationship with the ego. This representation has been defined in antropology for classical (offline)  social networks~\cite{dunbar1998social,hill2003social}, and it has been recently demonstrated that the same representation can also be adopted in OSN~\cite{arnaboldi2012analysis,PASSARELLA20122201}.

Epasto et al.~\cite{epasto2015ego} proposed a RS for friends recommendation based on the features derived from the co-occurrences of two nodes in different ego-networks.
In contrast to random walk-based measures, these features can be computed much more efficiently on very large graphs, by just analyzing selected neighbors of each node.
The authors evaluated the system on an anonymised snapshot of Google+ social network, and compared the proposed solution with two other unsupervised methods (i.e., common friends and Adamic-Adar) in terms of precision and recall. A 2-week test, during which the authors observed the social network evolution based on the different recommendations, demonstrated the effectiveness of their solution.

In general, the effectiveness of the proposed unsupervised methods highly depends on the social network structure used for their evaluation and they are not easy to compare~\cite{lichtenwalter2010new}.
On the other hand, supervised methods try to cope with this issue, by treating the link prediction in a social network as a classification problem and by defining the characteristics of the network in a data-driven way, i.e., by creating a vector of features for each pair of nodes, in which the set of features can include different heuristics and metrics.
To this aim, for each pair of nodes, a multidimensional vector is extracted and associated with a positive or negative label, based on the presence or absence of a link between the nodes.
Finally, a classifier is trained using the set of vectors as training data.
In this way, the classifier automatically learns the relevance of the features, and the learned model can be used to predict the creation of new links (i.e., the label) between any pair of nodes.

The following solutions are representative of this class of link prediction approaches.
Fire et al.~\cite{fire2011link} proposed a set of features related to different characteristics of the graph (e.g., nodes, edges, subgraphs, and multi-hop paths) in order to learn a classifier that identifies the missing links.
They used different datasets (e.g., Facebook, Flickr, YouTube, Academia, etc.) and by using a common framework they developed different classifiers for each dataset (e.g., decision tree C4.5, kNN, Naive Bayes, SVM, neural networks) in order to evaluate both single solutions and the combinations of some of them. Finally, they compared the prediction accuracy by using all the features or some limited sets.



Backstrom and Leskovec~\cite{backstrom2011supervised} proposed emph{Supervised Random Walks}, a supervised algorithm that combines the network structure with attributes about both nodes and edges in a unified link prediction algorithm.
Given a source node \emph{s} and training examples
about which nodes \emph{s} will connect to in the future, the algorithm learns a function that assigns a strength (i.e., random
walk transition probability) to each edge, providing higher scores to the possible new links of \emph{s}. 
This solution has been compared to both unsupervised methods (e.g., random walk + restart, Adamic-Adar, Common Friends) and other two supervised approaches (i.e., decision-tree e logistic regression) on heterogeneous datasets (i.e., arXiv and Facebook).
All the supervised methods outperformed the unsupervised ones in terms of precision and ranking metric AUC~\cite{hand2001simple}, and Supervised Random Walk performed better than the others of the same class.
We can thus observe that supervised methods generally performs better than unsupervised ones in friends and people recommendations, thanks to their ability to automatically learn the hidden relationships between social links and their features.

\subsubsection{Items recommendations}

Social context positively impacts also on generic items' recommendation (such as books, movies, or news), and several works have been presented in the literature exploiting most of the standard recommendation techniques.
The first category we consider is related to the neighborhood CF approach. These solutions use various strategies to select relevant users for the recommendations purpose.
For instance, authors in~\cite{victor2009comparative, zheng2008social, he2010social} considered only the connected friends of the target user and their items as the possible objects to recommend. Instead, in~\cite{he2010social} an evaluation between different strategies has been proposed. Specifically, the authors considered i) a Friends Average approach, which simply averages the ratings of the direct friends, (ii) a Weighted Friends, in which cosine similarity among friends is used to define links' weight, and iii) a traditional CF method without social context. All the social-aware recommendation tasks achieve better results than traditional CF in terms of prediction accuracy.

Other memory-based solutions exploit trust relations among users to obtain the set of relevant neighbors~\cite{golbeck2006generating, massa2007trust, jamali2009trustwalker}.
Generally, these solutions exploit different graphs' exploration techniques in order to propagate trust within the user-to-user network. In fact, in this case, the network is represents as a direct graph in which nodes represent the users and edges connect two users if they are friends or at least one of them declares a trust towards the other greater than 0. As a first step, the trust values among directly connected users are typically calculated leveraging on a user-to-user similarity measure (e.g., the cosine similarity calculated on the ratings of the two users)~\cite{tang2013social}. Then, these values of trust are propagated through the graph's edges in order to calculate the transitivity trust among nodes that are not directly connected to each other. For instance, \emph{TidalTrust}~\cite{golbeck2006generating} aims at estimating the trust values of a node (i.e., source) towards another one not directly connected (i.e., sink). In a first phase, it finds a path from the source to the sink while rating the nodes on the path, and then it aggregates the trust value backwards towards the source. As a result, TidalTrust assumes that: i) shorter propagation paths produce more accurate trust estimates, and ii) paths with higher trust values create better results in selecting neighbor nodes to be used for items recommendations. Therefore, TidalTrust selects as relevant neighbors only those nodes that are on the shortest path between the source user and the sink.
 \emph{MoleTrust}~\cite{massa2007trust} is another example, which uses a random walk to propagate trust values on the graph, and it defines a threshold on the length of the path between a source and a sink. Therefore, it limits the number of possible paths between a pair of users. 
 
In contrast with memory-based approaches, model-based recommender systems learn models that predict the user's interest for new items. As it happens for traditional RS, the matrix factorization (MF) represents the most used model-based technique in SRS.
The common idea of Social-aware MF solutions is that social connections mainly influence the users' preferences (i.e., the ratings). The preferences of two friends should be much more similar than those of two strangers, and for this reason these solutions typically associate a weight to the social relations, which indicates the strength of the social tie among the related users and it is usually calculated based on the ratings similarity.

Each of the MF solutions proposed in the literature differs from the others by the method used to include social and/or trust information into the factorization process.
For example, \emph{SoRec}~\cite{ma2008sorec}, instead of factorizing only the user-item matrix, as in traditional MF methods, it extends the factorization objective function performing a co-factorization both in the user-item matrix and the user-user social matrix.
Instead, other solutions (e.g., \emph{SocialMF}~\cite{jamali2010matrix} and \emph{Social Regularization}~\cite{ma2011recommender}) focus on the regularization term of the factorization objective function (see Section \ref{sec:mode-based_CF}), introducing new constraints related to the social factors.
Specifically SocialMF~\cite{jamali2010matrix} , by assuming that the behavior of a user is affected by the behavior of her direct neighbors,  introduces a new regularization term that forces the preference of a user to be closer to the average preference of the users in her social neighborhood.
On the other hand, Social Regularization~\cite{ma2011recommender} considers the fact that the target user's neighbors may have different tastes. For this reason it uses a pairwise regularization of the users' latent factors where the preferences' similarity of two connected users depends on their ratings' similarity. In order to evaluate Social Regularization, authors compared the performance of their solution with other three MF approaches: two traditional user-item MF (i.e., non-negative matrix factorization~\cite{lee1999learning}, and probabilistic matrix factorization~\cite{salakhutdinov2008bayesian}), and \emph{Social Trust Ensemble} (STE)~\cite{ma2009learning}, a social-aware MF solution were the missing ratings are predicted as a linear combination of ratings from the target user and her friends. By using the datasets extracted from two real-world OSNs (i.e., Douban~\footnote{http://www.douban.com} and Epinions~\footnote{http://www.epinions.com}), Social Regularization demonstrated to outperform the other approaches in terms of both MAE and RMSE metrics.
In addition, in the same work~\cite{ma2011recommender}, authors propose the use of two different similarity metrics for the pairwise regularization: the Pearson Correlation Coefficient and Vector Space Similarity~\cite{breese1998empirical}. However, the evaluation results show no significant differences between the two measures, but they both perform better than random similarity values, showing the importance of similarity function enhanced with social context information.

Recent works (i.e., ~\cite{sun2015recommender, reafee2016power}) use different regularization terms. In ~\cite{sun2015recommender} Sun et al. proposed \emph{RSboSN} (Recommender System based on Social Networks), which leverages on a clustering algorithm to identify groups of friends. RSboSN outperforms SoRec in terms of precision and recall, proving the effectiveness of their clustering-based regularization approach. Moreover, Reafee et al.~\cite{reafee2016power} proposed the  \emph{EISR} (Explicit and Implicit Social Relation Probabilistic Matrix Factorization ) algorithm, aimed at including both explicit and implicit social relations as regularization parameters. Explicit social relations are the direct friendships ties among users, while the implicit ones refer to not directly connected users and they are inferred using link prediction.
Compared with Social Regularization~\cite{ma2011recommender} and two diffusion-based algorithms (i.e., HeatS~\cite{zhou2010solving} and Hybrid~\cite{zhou2010solving}, described in Section~\ref{sec:network-based}), EISR shows a clear improvement in terms of prediction error (i.e., MAE and RMSE), demonstrating the effectiveness of taking into account also the implicit relationships in thesocial recommendations.

\subsection{Tag-based recommendations}
\label{sec:tag-based}

Over the past few years the use of special user-defined words, called tags, to categorize or describe web and OSN contents, gained a lot of popularity.
This user-driven phenomenon is known in the literature as \emph{folksonomoy}~\cite{mathes2004folksonomies} and it is a well-studied topic in both information retrieval and recommender systems fields~\cite{hotho2006information, zhang2011tag, milicevic2010social}.

An important aspect of folksonomies is that, differently from ontologies, no relationship between the terms is required a priori. On the contrary,  relationships are automatically built exploiting the tags created by the users and explicitly assigned to contents (e.g., hashtags assigned to tweets). Since tags are generally defined as keywords that can reflect a semantic meaning of the associated content, they represent an important feature in the content characterization. In addition, folksonomies dynamically adapt to changes in the users' vocabulary, further personalizing the user's interests, and they are independent from the type of item they describe .
For instance, the features used to represent the characteristics of a movie are typically very different from the features used to describe music files or books. Instead, tags can be seen as generic features that can be used to create multi-domain recommender systems~\cite{kim2014framework}.

However, since folksonomies have some drawbacks due to the user language. Synonyms, homonyms, polysemies, and different users' tagging behavior make the recommendation process difficult to perform in some cases, and undermine the use of simple tag matching (e.g., group items characterized by the same tag). To overcome these limitations, a novel family of recommender systems, called Tag-based Recommender Systems (TBRSs)~\cite{zhang2011tag, milicevic2010social}, have been proposed in the literature.

Formally, the folksonomy can be modeled in the following two ways. First, it is possible to use three distinct sets of elements: the set of users $U = \left\{u_1, \ldots, u_n \right\}$, the set of items $I = \left\{i_1, \ldots, u_m \right\}$, and the set of tags $T = \left\{t_1, \ldots, t_k \right\}$. Each binary relation between them can be described using adjacency matrices, $A^{UI}$, $A^{IT}$, and $A^{UT}$  for user-item, item-tag and user-tag relations respectively.
If the user $u_l$ has collected the item $i_s$, we set $a^{UI}_{ls} = 1$, otherwise $a^{UI}_{ls} = 0$. Similarly, we set $a^{IT}_{sq} = 1$ if the item $i_s$ has been tagged with $t_q$ and $a^{IT}_{sq} = 0$ otherwise. Furthermore, we set $a^{UT}_{lq} = 1$ if $u_l$ owns items tagged with $t_q$ and $a^{UT}_{lq} = 0$ otherwise.
Alternatively, the folksonomy can be modeled as a third-order tensor taking into account only complete ternaries $Y = (u,i,t)$. Specifically, the tensor's component is set $Y = 1$ if the ternary exists in the folksonomy, and $Y = 0$ otherwise.

\begin{figure}[t]
\begin{subfigure}{.5\textwidth}
\centering
\includegraphics[width=.7\linewidth]{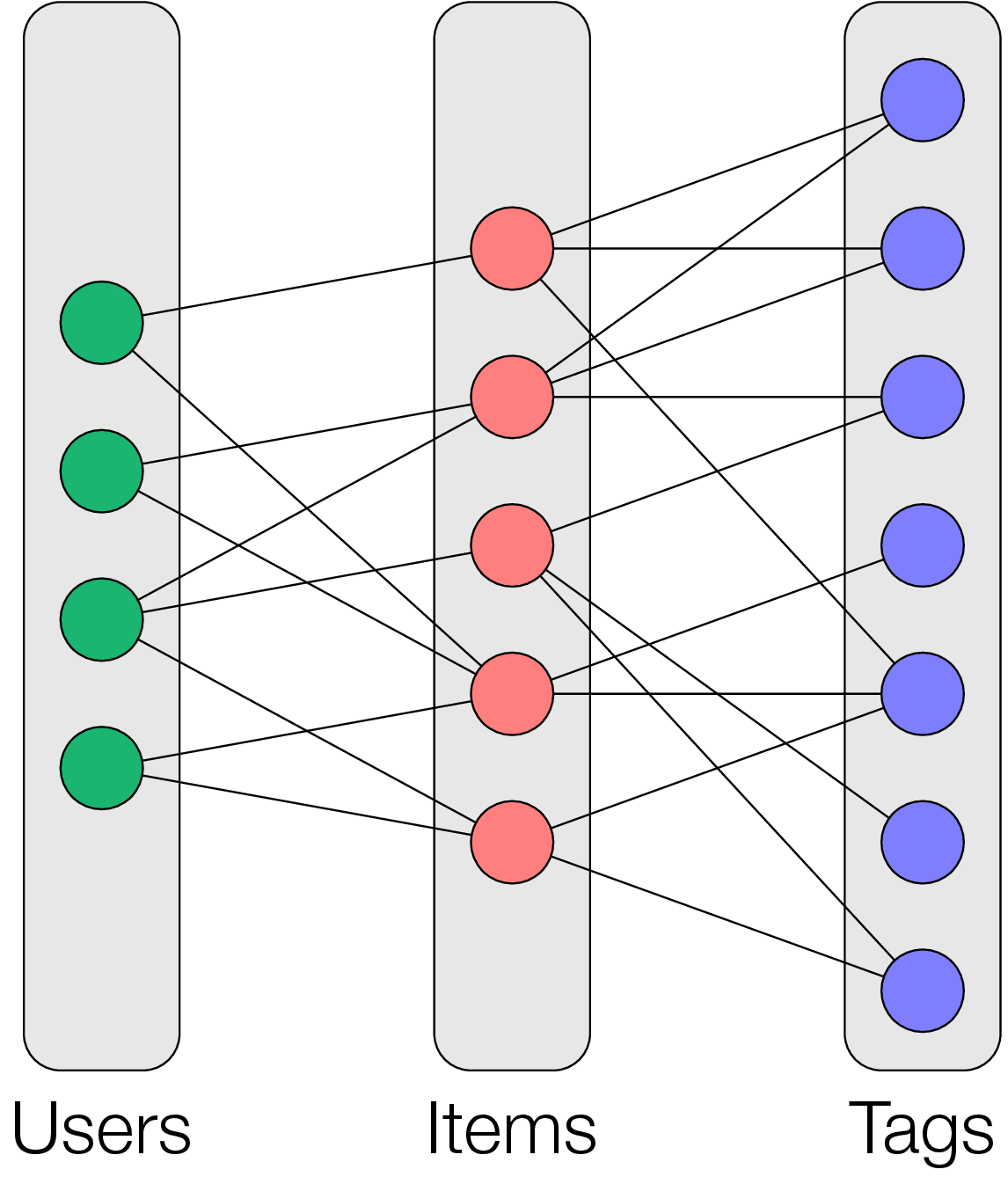}
\caption{}
\label{fig:tag-based-net}
\end{subfigure}%
\begin{subfigure}{.5\textwidth}
\centering
\includegraphics[width=.65\linewidth]{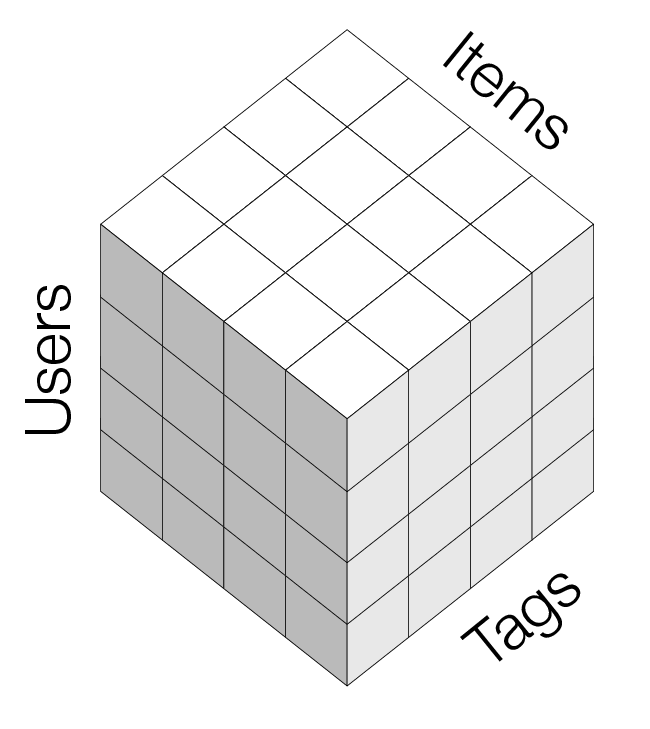}
\caption{}
\label{fig:tag-based-tensor}
\end{subfigure}
\caption{Representation of a folksonomy with (a) a graph, and (b) a tensor.}
\label{fig:fig}
\end{figure}

A variety of different approaches have been proposed in the literature for tag-based RS. Xu et al.~\cite{xu2006towards} and Sigurbj{\"o}rnsson et al.~\cite{sigurbjornsson2008flickr} investigated the tag co-occurrence frequency to support the user tagging process.
Specifically, given a specific item (e.g., a photo in Flickr), the system recommends to its users the most relevant tags to associate with it.
The underlying idea is that, if the co-occurrence frequency of two tags is high, they could be closely related. On the contrary, if two tags are not related at all, their co-occurrence frequency should be very low.

Several works extend CF towards tag-based RS. To this aim, the user-item-tag ternary is typically reduced to two-dimensional matrices. Firan et al. ~\cite{firan2007benefit} leverage on the user-tags projection matrix to compute a ranked list of tags, then a list of recommended items is extracted according to those tags. Instead, Giu et al.~\cite{guy2010social} select neighbors users by using a similarity measure based on common tags and items, while~\cite{peng2010collaborative} exploits the co-occurrence of tags.


In order to overcome the issues related to the folksonomies (e.g., synonyms and polysemies) and to improve the performance of CF, ~\cite{shepitsen2008personalized} and \cite{gemmell2008personalizing} use hierarchical tags clustering techniques, while~\cite{liang2010connecting} weights each tag based on the relationships between users, items, and tags. Recently, a hybrid framework for both item and tag recommendations has been proposed in~\cite{kim2014framework}. Authors use text analysis techniques (e.g., bigrams) to solve the ambiguity problem of folksonomies and perform tag recommendations. In addition, they exploit temporal information and trust relationships among the users to provide item recommendations using a CF approach.

Modeling the folksonomy as a third-order tensor, some researchers proposed recommender systems based on tensor factorization techniques (see Section~\ref{sec:cars}). Specifically, in~\cite{symeonidis2008tag} the recommendation is performed by the product of three low rank matrices (i.e., the latent components of the users, items, and tags) and a low rank core tensor produced by the Higher-Order Singular Value Decomposition (HOSVD)~\cite{de2000multilinear}.
Rendle et al.~\cite{rendle2009learning} proposed a different learning approach for tensor factorization models called Ranking with Tensor Factorization (RTF), which optimizes the model parameters for the ranking statistic AUC~\cite{hand2001simple}.

A natural way to represent the folksonomies is to use hypergraphs. According to this representation, nodes of the graph represent the folksonomy entities (i.e., users, items, and tags), and the relationships between them are modeled with edges among nodes. Consequently, a number of researchers proposed different graph-based solutions for tag-based RS. These approaches exploit the structure of the folksonomy hypergraph in order to provide tags or items recommendations to the user. Hotho et al.~\cite{hotho2006information} proposed FolkRank, a graph-based ranking algorithm inspired by PageRank~\cite{brin2012reprint}. Here, the underlying idea is that an item tagged with ``important'' tags by ``important'' users becomes ``important'' by itself. Essentially, FolkRank identifies the most ``important'' tags (i.e., the most visited tags during the random walk) for a specific user by performing a biased random walk with restart on the folksonomy graph. 

Other solutions refer to the diffusion-based approach described in Section~\ref{sec:network-based}. Zhang et al.~\cite{zhang2010personalized} were the first to propose a diffusion-based recommender system based on the tripartite user-item-tag graph. More specifically, they apply the diffusion process proposed in~\cite{zhou2007bipartite} on both the user-item and item-tag bipartite networks and then, they combine the two results to provide the recommendations to the user.
Their results show that the integration of tags can enhance the accuracy of the recommendations with respect to previous diffusion-based algorithms.
However, due to their diffusion process, the solution proposed in~\cite{zhang2010personalized} and other similar works tend to suggest the most popular items in the folksonomy network, penalizing the personalization of the recommendations~\cite{zhou2010solving}.

To this aim, we recently proposed PLIERS (Popularity-based ItEm Recommender System), a tag-based RS designed both for bipartite~\cite{arnaboldi2016pliers} and tripartite graphs~\cite{arnaboldi2017personalized} .
In order to provide even more personalized items to the users, PLIERS assumes that the popularity of an item/tag (i.e., the degree of the node) can be related to its semantic.
Therefore, a very popular item or tag can semantically relate to a more ``generic'' topic compared to a less popular item/tag that, instead, describes a more ``specific'' topic.
For instance, any content related to the football club Millwall can be tagged with both tags \emph{``Millwall''} and \emph{``Football''}, but the opposite is not always true: all content concerning football will not always be tagged with \emph{``Millwall''}. According to this assumption, we can therefore say that the tag \emph{``Football''} refers to a more generic topic than that referred by the tag \emph{``Millwall''}.
Users interested in the Millwall football club, but not connected to items tagged with \emph{``Football''}, are clearly not interested in all the items tagged with the latter tag, as these could contain information about other football clubs.
In this ways, PLIERS solves the dilemma of the choice between popular or non-popular items in the network in a more natural way than the other diffusion-based algorithms. PLIERS does not without require any tuning parameter, and it ensures that the popularity of recommended items is always comparable with the popularity of items already adopted by the users.
Compared with other diffusion-based solutions, PLIERS shows comparable performance in terms of precision and recall, but providing better novelty in the recommendations.

A very recent and promising approach to build tag-based recommender systems is represented by the use of deep learning to discover latent features from the tag space. The system proposed in ~\cite{zuo2016tag} uses a sparse autoencoder~\cite{hinton2006reducing} to extract a set of dense latent features from the user-tag matrix. Based on the extracted features, users' profiles are updated and the traditional CF is used to recommend items or tags.
On the other hand, ~\cite{ijcai2017-446} proposes a pure deep learning model to provide item recommendation to the user.
Here, two neural networks map the tag-based user and item profiles to an abstract deep feature space. Then, the relevance of the specific item for the user is calculated based on the similarity between the deep representation of the two profiles.
Authors compared their solution with three different approaches: a hierarchical clustering model~\cite{shepitsen2008personalized}, CF and the autoencoder-based model proposed in~\cite{zuo2016tag}. With respect to these solutions, experimental results show that the proposed approach significantly improves the effectiveness of the recommendations in terms of both precision and recall.

\subsection{Location-aware recommendations}
\label{sec:lbsn}

The recent advances in location-acquisition techniques, and the wide spread use of GPS-enabled smartphones, led to the creation of the so-called Location-Based Social Networks (LBSN), such as Foursquare~\footnote{foursquare.com} or Yelp~\footnote{www.yelp.com}.
These systems allow users to tag, rate, and describe the locations they visit, in order to aid the discovery of unknown points of interests (POIs), like restaurants or shops, or to enable new social relations among users~\cite{quercia2010recommending}.

\begin{figure}[t]
\centering
\includegraphics[width=0.8\textwidth]{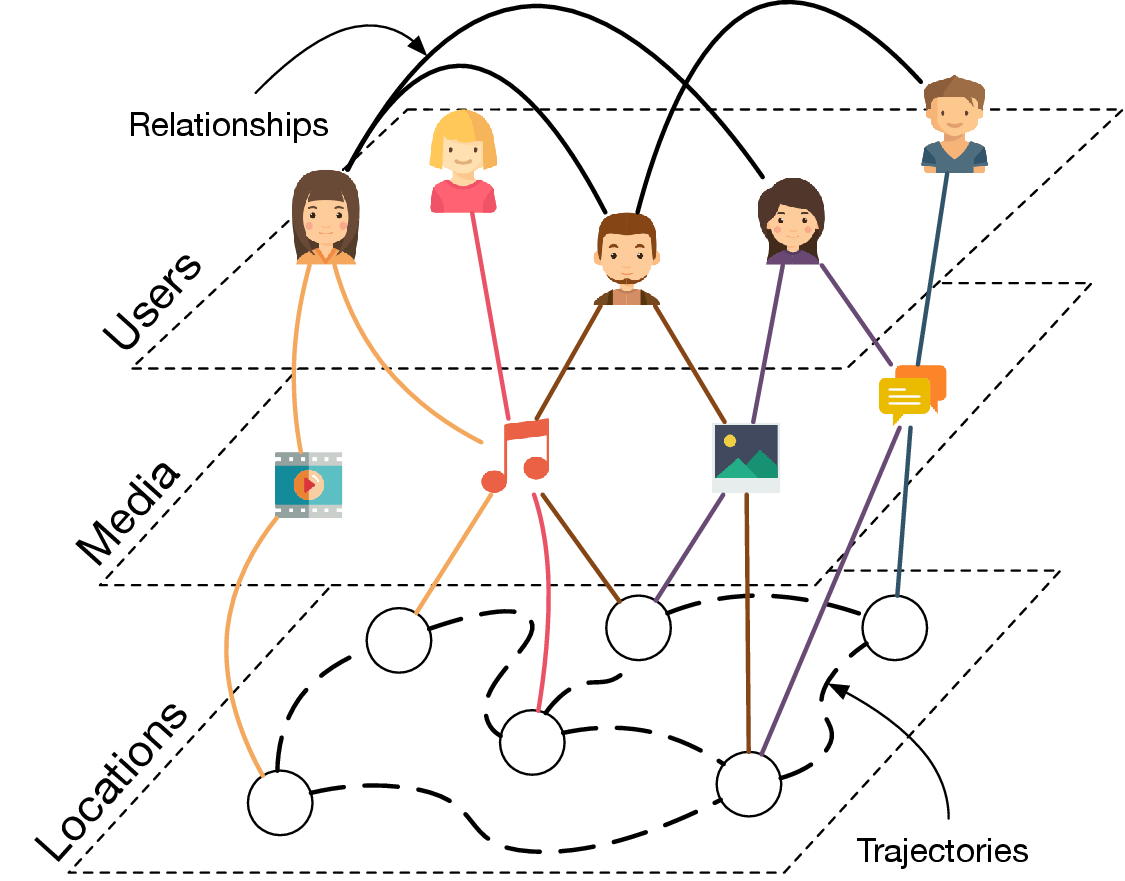}
\caption{Data in Location-based Social Networks.}
\label{fig:loc_based_dimensions}
\end{figure}

According to Bao et al.~\cite{bao2015recommendations}, the location is one of the most important information to describe the user's context. The addition of spatial information in social networking systems is able to bridge, at least partially, the gap between cyber and physical social relationships among users. Figure~\ref{fig:loc_based_dimensions} depicts the main information modeled by a typical LBSN. In the upper layer, users and their social relationships are represented. These social relationships can be explicitly defined by the users (e.g., friends or followers) or they can be inferred from their activities on the OSN. The middle layer depicts the social media content generated by the users. The nature of these content can be very different; for instance, a user can share a check-in at a public place, a geo-tagged photos, a review of a restaurant, or the workout tracked by a fitness application.
As depicted in the lower layer, each content should be associated with one or more geographical locations.


Given the multi-dimensional nature of the LBSNs, SRS applied to these specific OSN must be able to handle a multitude of heterogeneous data. They can be divided in four different categories based on their recommendation task: i) location, ii) user, iii) activity, and iv) social-media.
The first category mainly focuses on suggesting new locations to the user based on her preferences.
The objective of the recommendations can be both single POIs or a sequence of locations, such as the best route to take or the sequence of attractions to visit during a trip.
To recommend a single place, both content-based and collaborative filtering techniques have been exploited. In the former case, the solutions proposed in the literature, e.g., ~\cite{kodama2009skyline}, ~\cite{park2007location}, ~\cite{ramaswamy2009caesar}, typically suggest new locations by matching the user's profile with the places' meta-data (e.g., text description, categories, or tags). These solutions inherit the main advantages and drawbacks of the content-based approach: they are effective in mitigating the cold-start problem but, on the other hand, they may suffer from the recommendation quality issue. In fact, taking into account just the preferences of the target user and ignoring the opinions coming from other users, the system may recommend a matching place with poor quality from the social opinion standpoint.
In order to overcome this problem, a large number of works exploit the users' location histories and CF models. Leveraging on the other users' past actions (e.g., reviews or check-ins), these location-based RSs are able to improve the quality of the recommendations by ignoring poorly-reviewed locations that otherwise match the target user's profile.
As we argued in Section~\ref{sec:collaborative_filtering}, the crucial point in the CF's recommendation process is represented by the selection of neighbors users (or items).
The early solutions define their neighbors selection according to the first law of geography of Tobler~\cite{tobler1970computer}, which claims that \emph{``everything is related to everything else, but near things are more related than distant things''}.
For instance, the work in ~\cite{horozov2006using} considers only individuals who live near the location from which the target user has made the query, while in ~\cite{chow2010towards} candidate POIs are pruned by using a spatial range around the user's current location.
In addition to the spatial distance between places and users, several works proposed to use a different kind of information in order to further improve the quality of the recommendations, for example, \cite{shi2011personalized} proposes a category-regularized matrix factorization model, taking into account the category of the places visited by the user in the past (e.g., ``Italian restaurant'' or ``Irish pub'').

Sentiment analysis can also be used in order to build a more fine-grained model of the user's preferences~\cite{moshfeghi2011handling, levi2012finding}. Yang et al. \cite{yang2013fine} exploit the user's comments left in the check-ins in a tensor factorization framework to build a personalized location ranking system, while the system proposed in ~\cite{levi2012finding} considers as neighbors not just the users with similar preferences, but also those with comparable needs. Specifically, it clusters the places' reviews according to different aspects, such as the
the topics covered in the text (e.g., service, food, facilities), the intent and the nationality of the author.
In a recent work, Wang et al.~\cite{wang2017location} face the problem of POIs recommendation considering the \emph{user interest drift}~\cite{yin2016adapting}. According to this phenomenon, users tend to have different interests when they travel out of their hometown. To this aim, the authors propose a model to simulate the decision-making process of users' check-in behaviors, both in hometown and out-of-town areas, using both geographical clustering and topic modeling (i.e., Latent Dirichlet Allocation - LDA~\cite{blei2003latent}) on users' check-ins and reviews.

Furthermore, other works explore the influence of social aspects in location-based recommendations. Assuming that friends share many more preferences than strangers, Ye et al.~\cite{ye2010location} limit the choice of neighbors to the user's friends only.
They extend their work in~\cite{ye2011exploiting} by selecting candidate neighbors taking into account the user preferences extracted from check-ins histories and social connections, but also the geographical distance between users and candidate places. They also perform a comparative evaluation among different CF and random walk-based approaches, proving that the geographical factor has more impact on the accuracy of the recommendations than personal social relations. However, social links better address the cold-start problem because social friends may provide potentially relevant POIs to new users without location histories.
Therefore, the combination of different aspects, such as geographical influence, social relations and user preferences, ensures to obtain the highest performance in recommending individual locations~\cite{gasparetti2017personalization}.

As previously described, SRS can provides people/friend recommendations mainly exploiting the underlying social structure and user interaction patterns~\cite{wiese2011you, backstrom2011supervised, epasto2015ego}.
However, it has been proven that location has significant correlation with users' social behaviors in their real life, such as friendship relationships~\cite{liben2005geographic, cranshaw2010bridging}.
Therefore, LBSNs can provide also a new way to make people/friend recommendations taking into account the users' location histories.
For example, ~\cite{zheng2009mining} and ~\cite{ying2011user} focus on finding popular users in LBSNs. They consider as ``popular'' the users with more knowledge about the locations (e.g., those who made several check-ins) and they assume that this kind of users are able to provide high quality location recommendations.
A large body of research (e.g.,~\cite{cho2011friendship, xiao2010finding, zheng2011recommending, yu2011geo, scellato2011exploiting}) leverages on users' location histories to improve the effectiveness of friend recommendations. Here, the main idea is that the history of visited places can reveal preferences, and thus people with similar location histories may have similar preferences and may become friends more likely.
Scellato et al.~\cite{scellato2011exploiting} propose a supervised learning framework that exploits users' visited locations to predict new links among \emph{place-friends}, i.e., users who visit the same places, and \emph{friends-of-friends}. Their results show how the inclusion of information about places, and related user activities, offers high link prediction probability for the friend recommendation task.
Another example is represented by~\cite{xiao2010finding}, which estimates the similarity between users according to the semantic of their visited locations (e.g., restaurants or cinemas). This allows the system to recommend connections between users who have different geographic behaviors (e.g., living in different cities), but share similar semantic behaviors , e.g., they frequently visit the same types of locations.

Leveraging on the data shared by users in LBSNs, it is also possible to extract useful knowledge about locations and user activities. For instance, users may be interested in knowing which activities (e.g., dining, shopping, watching movies/shows, enjoying sports/exercises) can be practiced in a given place.
Activity recommendations can be performed by exploiting both the meta-data associated with the locations (e.g., tags)~\cite{pozdnoukhov2011space} and the users' location histories~\cite{yin2011geographical}.
On the other hand, other works provide activity recommendations by exploiting information from all the users with collaborative-based approaches. For instance, Zheng et al.~\cite{zheng2010collaborative} and Symeonidis et al.~\cite{symeonidis2011geo} proposed the use of a 3-order tensor to charcterize each user check-in operation. the tensor includes the current user's location, the performed activity and a rating on that activity.  They use  factorization methods on these tensors in order to provide location-specific activity recommendations. 

The last type of recommendations that can benefit from the use of geographical information is the content recommendation, with specific reference to social media contents. In this case, the system provides users with suggestions of photos, videos, or other web content they might like, which have been shared by other users through a OSN. Several works exploit the histories of users location to improve the quality of social media recommendations. For instance, ~\cite{kawakubo2011geovisualrank} proposes a picture ranking algorithm that exploits the locations in order to improve the relevance of search results; ~\cite{silva2011tag} presents a method to suggest the most relevant tags for georeferenced photos, and ~\cite{sandholm2011real} builds a location-based RS aimed at increasing the diversity of recommendations, while mitigating popularity bias of web contents.
In a recent work, P\'{a}lovics et al.~\cite{palovics2017location}
address the problem of recommending highly volatile items, as locations or events characterized by strict time constraints (e.g., a concert or an expo). Specifically, they use both online machine learning and matrix factorization techniques to recommend to their users the most relevant hashtags for a given geo-referenced tweet at a given time. Their solution performed much better than online matrix factorization~\cite{palovics2014exploiting} and content-based~\cite{harvey2015long} methods, which justifies the importance of temporal and geographic information in the social-media recommendation task.

\section{Recommender Systems for Mobile Social Networks}
\label{sec:msn}


In Mobile Social networks (MSN), typically, a mobile device should be able to autonomously share information with other devices in proximity in order to discover useful contents for its user.
In this context, RS represent a useful tool to improve the content dissemination in MSN, by proactively suggesting to the users the information discovered in the nearby.
However, given the distributed nature of MSN, the solutions proposed for OSN (discussed in Section~\ref{sec:osn}) cannot be simply extended to this new scenario. In fact, RS for OSN generally rely on standard client/server centralized models, where the recommendation engine runs on the server side (or a cloud-based infrastructure) and processes the requests coming from fixed and mobile clients.
In addition, the recommendation task in MSN significantly differs from that defined for the online environment.
As discussed in Section~\ref{sec:reccomendation_task}, RS for OSN typically focus on learning a predicting model for missing values in the ratings matrix.
This matrix can be seen as a systems' global knowledge about both the available items and users' preferences. Thus, the recommendation engine learns a single global model which will be used to perform the recommendations for all its users.
On the contrary, in MSN each device may be aware of just a (small) fraction of the global information. This local knowledge is initially related only to information about the local user, then it grows up with those exchanged with encountered users and devices, through D2D communications.
Therefore, in MSN each device should learn a recommendation model focused on the preferences of its local user.
	
The first approaches proposed in the literature for RS in distributed environments refer to the Peer-to-Peer (P2P) paradigm.
In a P2P architecture, RS are fully distributed over several nodes, which act as client and server at the same time~\cite{androutsellis2004survey}.
The solutions proposed in this context (e.g.,~\cite{han2004scalable, miller2004pocketlens, kim2008user}) are typically based on Neighborhood CF. Specifically, the user-item ratings are maintained in a distributed way on the P2P infrastructure and, when a peer has to make a prediction for its local user, it exploits P2P lookup methods to find and retrieve the relevant ratings necessary to locally calculate the rating for the target item.

By shifting the recommendation process to the client side, P2P solutions are able to alleviate some critical points of centralized methods; for example, the possible bottleneck represented by the server, or security and privacy issues that can arise when data is handled by a single organization ~\cite{zhao2016effect}\cite{Lam2006}.
However, by using standard P2P systems, peers should be able to communicate to each other by using infrastructured networks (i.e., accessing the Internet), in order to have a constant access to the entire knowledge of the network. This is not always possible in MSN and, in the last few years, some RS explicitly tailored for MSN and opportunistic networks has been presented in the literature.

We classify the proposed solutions based on the RS method used: i) collaborative filtering, and ii) tag-based, previously described in Sections~\ref{sec:collaborative_filtering} and \ref{sec:tag-based}, respectively.
Hereafter we present the main solutions belonging to the two categories.

\subsection{Collaborative Filtering}

The first set of RS proposed in the field of MSN use CF.
In this case, the recommendation task is performed on each local device, and it is based on the users' ratings exchanged among nodes via D2D communications.

The proposed solutions differ in two main aspects: (i)
the metrics used to measure the similarity between users,
and/or (ii) the heuristics implemented to reduce the computational complexity of CF in a mobile environment.

The simplest similarity measures refer to the rating-based metrics used in standard CF (e.g., the Pearson Coefficient discussed in Equation~\ref{eq:user-pearson}), while other solutions, to further improve the quality of the recommendations,  take into account additional context information.
	
\subsubsection{OppCF}
De Spindler et al.~\cite{de2007collaborative} are among the first in the literature who investigated the use of RS in opportunistic environments. Specifically, they used a distributed user-based CF approach to predict ratings based on shared information among co-located mobile devices.
They present a preliminary idea of ``\emph{social context}'' shared among mobile devices during their opportunistic contacts and aimed at improving the recommendation process. In this case the social context is identified by the co-location of users related to a specific event (e.g., attending a conference or a concert). This condition is the basic assumption of the authors to identify similar users, as those who are physically co-located in a specific time interval. 
The authors do not consider the possibility to exploit also the occasional co-locations of users as potential source of interesting contents.  
This pioneering solution has never been evaluated, thus its accuracy cannot be assessed.

\subsubsection{MobHinter}
MobHinter ~\cite{schifanella2008mobhinter} presents two main contributions: (i) the definition of \emph{Affinity networks} aimed at identifying similar users, and (ii) an epidemic protocol designed to distribute and exchange users' ratings among mobile devices in ad-hoc networks. The authors focus on the concept of {Affinity on ratings}, by calculating the percentage of similar ratings provided by two users on a common set of items, and by assigning a threshold value to identify similar users. Therefore, to apply CF method based on users' affinities, MobHinter proposes an epidemic protocol to exchange a set of ratings information through ad hoc communications: (i) the list of ratings of the local node, (ii) the list of ratings of its current neighbors and (iii) the ratings of the nodes encountered in the past. Then, each node is able to locally compute its affinity network, and to apply standard CF methods to predict missing ratings and provide the recommendation. The authors proposed also optimized versions of the protocol, by exchanging only the list of ratings of the local node or by excluding those of the nodes encountered in the past.

Authors evaluated their proposal using the MovieLens dataset and simulating a fully connected scenario (to estimate the optimal behavior) and an ad hoc network scenario based on a random mobility model. In the former scenario, they compared their model with other user-based CF methods, evaluating the prediction accuracy (i.e., MAE and RMSE). Then, they performed a simulation to evaluate the number of random meetings needed on average by users to approximate the reference prediction accuracy obtained in the fully connected scenario. The authors showed that by exchanging all the ratings information known by each user, the system approximates the reference accuracy with a limited number of meetings. However, the use of a random mobility model is not realistic to simulate an opportunistic networking scenario as largely demonstrated in the literature~\cite{karamshuk2011human}.
	
\subsubsection{diffeRS}
Del Prete and Capra~\cite{del2010differs} proposed an interesting approach to reduce the complexity of CF in MSNs.
They proposed \emph{diffeRS}~\cite{del2010differs}, a decentralised RS aimed at locally classify the user as \emph{mass-like minded} or \emph{individual}, based on the increasing knowledge about the other encountered users and their profiles.
In fact, also in this case, mobile devices exchange the users profile through D2D communications (by exploiting Bluetooth technology, in this case).
For each recommendation task, diffeRS evaluates the average deviation of the local user's profile with respect to the preferences expressed by the local community as a whole (i.e., the average ratings).
If the local user is identified as \emph{mass-like minded}, the recommendations are simply based on the average of the community preferences, thus reducing at the minimum the computation on the mobile device.
On the other hand, if the local user is an \emph{individual}, a user-based CF approach is applied, considering only other individual users that differ from the community's preferences in the same proportion of the local user.
To this aim, the system selects only those neighbors that share with the local user at least one rated item, and it computes the difference among their co-ratings to define a measure of their similarity with respect to their deviation from the average of the community preferences.
In this way, diffeRS further redices the sparsity of the rating matrix used in CF.
Therefore, diffeRS reduces the recommendation complexity for both user profiles, by exchanging a limited amount of information during the opportunistic contacts.

The authors evaluated diffeRS both in a fully connected scenario and in an opportunistic one by using two datasets: MovieLens~\footnote{https://grouplens.org/datasets/movielens/} to model users' ratings, and MIT Reality Mining~\cite{mit-reality-20050701} to model users' mobility.
In the fully connected scenario, the authors evaluated the prediction accuracy in terms of MAE and they demonstrated that the proposed correlation measure, divided between \emph{mass-like minded} and \emph{individual} performs better than the standard Pearson correlation.

In the opportunistic scenario, they randomly mapped 100 users from the MovieLens rating dataset to the 100 users of the MIT Reality Mining dataset. Predictions are computed at regular intervals of time, comparing accuracy (i.e., MAE) and coverage of diffeRS with respect to CF based on Pearson correlation. 
The obtained results show that diffeRS always outperforms the user-based CF. 

\subsubsection{locPref}
In a recent work, Zhao et al.~\cite{zhao2016robust} proposed the use of CF to recommend privacy preferences depending on the user location.
They refer to a Location Sharing Service (LSS) scenario, where users can share their current location using the ``check-in''' functionality of social network applications.
The declaration of a user to be in some specific location (e.g., a clinic or religious sites) may be considered as a sensitive information, and people usually may not wish to share them with the others~\cite{beresford2003location}.
LSS can autonomously share location information, following the general privacy settings of the user. LocPref is aimed at recommending the user with specific privacy settings for sharing her current location through the LSS. To this aim, it leverages on the assumption that people who visited the same location can share similar privacy preferences. Therefore, users' mobile devices can exchange location-privacy preferences through opportunistic communications and locally compute their recommendations.

The authors also proposed a simple but elegant mechanism to reduce the vulnerability of CF to \emph{shilling attack}~\cite{lam2004shilling}: the tentative of malicious users (i.e., attackers) to bias the recommendations by giving false information and/or ratings~\cite{burke2015robust}. The attackers can create fake profiles (also called \emph{shill profiles}) in order to inject modified data into the recommendation process.
For instance, in the context of LSSs, a business owner may force the customers to share their location when they visit her shops in order to increment their visibility on social media. In the same way, she may want to prevent the customers' check-ins at competitors' shops to reduce their popularity.

In order to prevent shilling attacks, authors proposed a reputation scheme based on the encounter frequency of nodes. Every time a node receives a user's profile from another node, it increments a counter associated with that specific profile and considers it as the reputation level of the encountered user. Once the system needs to perform a recommendation, it uses only those profiles whose counter is greater than the average reputation of all the profiles in the local cache (i.e., the average number of encounters of all nodes in the cache).
An attacker usually creates multiple fake profiles, but in an opportunistic environment, it can share just one of them during a contact. For this reason, this simple mechanism is able to associate a lower reputation with the shill profiles than those calculated for real ones.

To evaluate the proposed solution, authors implemented in locPref two different schemes for the opportunistic data exchange: (i) nodes exchange only the local user profile (in terms of privacy settings) at each opportunistic contact (\emph{D-Ind}); and (ii) nodes exchange both the local profile and those previously received by the others (\emph{D-Set}).
In order to obtain a reference evaluation metric, they simulated the system in a fully connected environment and then in a mobile scenario. They exploited the \emph{st\_andrews/locshare} dataset available in CRAWDAD~\cite{st_andrews-locshare-20111012}, which contains the location-privacy preferences of 40 users collected in the city of St Andrews (UK), and they simulated the user mobility patterns based on St Andrews road map and selected POIs in the city.

Simulations results showed a comparison between \emph{D-Ind} and \emph{D-Set} versions of locPref, highlighting the time needed to reach the same accuracy of the fully connected scenario (comparable with a centralized solution) and the advantages of exploiting the exchange of additional profile information to compute the recommendations.
In addition, the authors proved the efficiency of their reputation-based mechanism, which allows to prevent the recommender system from being abused by shilling attacks.

\subsection{Exploiting tags}

As we described in Section~\ref{sec:tag-based}, exploiting user-defined tags (i.e., folksonomies) is an effective method to improve the quality of the recommendations.
Tags can be also used in MSNs to further characterize contents even from a semantic point of view.
However, in order to implement tag-based MSN solutions, we need a mechanism which is able to infer the relations among different tags using just the user's preferences and the limited information obtained by other nodes in proximity.
In the following, we present in detail the two solutions proposed, so far, in the literature, which exploit tags in MSNs: ICe-Habit~\cite{lo2010folksonomy} and p-PLIERS~\cite{arnaboldi2017personalized}.




\subsubsection{Information-Centric Habit (ICe-Habit)}

To the best of authors' knowledge, ICe-Habit~\cite{lo2010folksonomy} is the first approach proposed in the literature that exploits folksonomies to improve the performance of a content dissemination protocol (i.e., Habit~\cite{mashhadi2009habit} ) for opportunistic networks.

In ICe-Habit, each user's profile is characterized by a vector of tags used in the past by the user to describe her contents.
User profiles are exchanged by nodes during opportunistic contacts.
In this way, each node can locally build a tag co-occurrence matrix $M$. Each entry of the matrix, $M[i,j]$ represents the number of items characterized by the co-occurrence of tags ($t_i$,$t_j$).
Therefore, $M$ is used to select contents to be forwarded to potentially interested users in opportunistic networks.

Whenever a user creates a new content, she associates to it a set of tags $T = \left\{ t_1, \cdots, t_n \right\}$.
Then, the system performs the \emph{tag-expansion} procedure to further enrich the set of tags $T$.
Specifically, the system queries the matrix $M$ to extract the $k$ tags (i.e., $T_k$) with the highest value of co-occurrence with those specified in $T$.
Then, the expansion set of tags $T' = T \cup T_k$ is associated to the new content.
Finally, the item is opportunistically forwarded to the potentially interested users, i.e., each user $x$ whose profile $T_x$ contains at least one of the tags in $T'$.

The authors evaluated ICe-Habit, via simulation, in an opportunistic scenario.
To this aim, they used the MIT Reality Mining datasets, which contains the contact traces of 96 nodes.
In addition, the MovieLens dataset has been used to create a user's profile for each node.
Specifically, each profile contains the set of tags the user has created to tag movies in the dataset.
Furthermore, for each $\left<user_i, tag_j \right>$ pair, the MovieLens dataset contains also the timestamp information in which the $tag_j$ has been created by the $user_i$.
The authors used the timestamp information to map the creation of new items on the simulated time. 
In order to measure the performance of the proposed solution, authors used the following approach: when a node creates a new content, it does not include all the tags it would normally associate to it, but it randomly drops 50\% of the item's tags. In this way, authors measured the \emph{Tag Recovery} and \emph{Destination Recovery}, which are, respectively, the percentage of the dropped tags and correct destination reached by each message that could be recovered by the tag expansion method.

Although the results demonstrate that the tag expansion improves the content dissemination's performance in a MSN, they also prove that this approach is not able to correctly infer the semantic relations between different tags.
In fact, considering just the tags' co-occurrences, nodes could receive more items than those really interesting for them because users may not be interested in the topics represented by the expanded tags.

\subsubsection{Pervasive PLIERS (p-PLIERS)}

Recently, we proposed a novel framework for the content discovery and evaluation in MSNs called \emph{pervasive PopuLarity-based ItEm Recommender System} (\emph{p-PLIERS})~\cite{arnaboldi2017personalized}. It can be used both as standalone RS for MSNs, and as a useful support tool for content dissemination or routing protocols in fully distributed environments (e.g., opportunistic networks).

Differently from ICe-Habit~\cite{lo2010folksonomy}, p-PLIERS does not leverage just on the tags' co-occurences. It exploits all the folksonomy's information, i.e., the relations among users, shared or created items, and tags used to semantically describe the contents.
This information is locally maintained by each node, and modeled as a tripartite users-items-tags graph, which is called \emph{Local Knowledge Graph} (\emph{LKG}).
The LKG of each node merges the information about the items created or downloaded by its local user, with the local knowledge of other encountered nodes, obtained through opportunistic communications.
Specifically, when a node creates a new item, it updates its LKG by inserting the relation between its user entity, the generated item and the related tags. 
Similarly, when a node encounters another node in the network, the two nodes exchange their LKGs, and locally integrate them with the received information.

Once each node has updated its local LKG with the information coming from the other node, p-PLIERS implements the PLIERS recommender system~\cite{arnaboldi2016pliers} (described in Section~\ref{sec:tag-based}) to evaluate the relevance of the new available items with respect to the interests of its local user.


As a first set of experiments, we evaluated the accuracy of PLIERS recommendations in a centralized scenario with respect to other solutions proposed for MSNs: the Tag Expansion mechanism exploited by ICe-Habit~\cite{lo2010folksonomy} and user-based CF used in similar works (e.g., MobHinter~\cite{schifanella2008mobhinter} or diffeRs~\cite{del2010differs}).
To this aim, we extracted a real dataset from Twitter using the Twitter Streaming API~\footnote{dev.twitter.com/streaming/overview}. Specifically, we downloaded the tweets generated during a big event (i.e., the World Food Day at Expo 2015) in the urban area of Milan, and we built a tripartite graph composed by more than 5000 users, 2946 tweets, and 3202 hashtags.
This represents a realistic folksonomy user-item-tag graph of online tagged contents related to a popular event.
To assess the accuracy of the recommendations, we performed a \emph{link prediction} task on a tripartite graph: we removed one link from each user connected at least to 5 items with popularity greater than one (i.e., the number of connected users).
Then, we calculated the performance of PLIERS, Tag Expansion and user-based CF in terms of precision and recall, based on the number of links that are included in the recommendations of each algorithm (i.e., the ``recovered links'').
In the centralized scenario, PLIERS outperformed the two reference algorithms for both considered measures.
Specifically, it reached a precision score up to eight times higher than that obtained by Tag Expansion and a score 40\% higher than CF.
With regard to the Recall measure, PLIERS obtained a score around 50\% higher than Tag Expansion and around 30\% higher than CF.
These results indicate that PLIERS, exploiting all the information contained in the folksonomy graph (i.e., user-item-tag relationships), is able to obtain better results than the other solutions proposed for MSNs, which consider only part of the user-item or item-tag relationships.

\begin{figure}[t]
	\centering
	\includegraphics[width=0.95\textwidth]{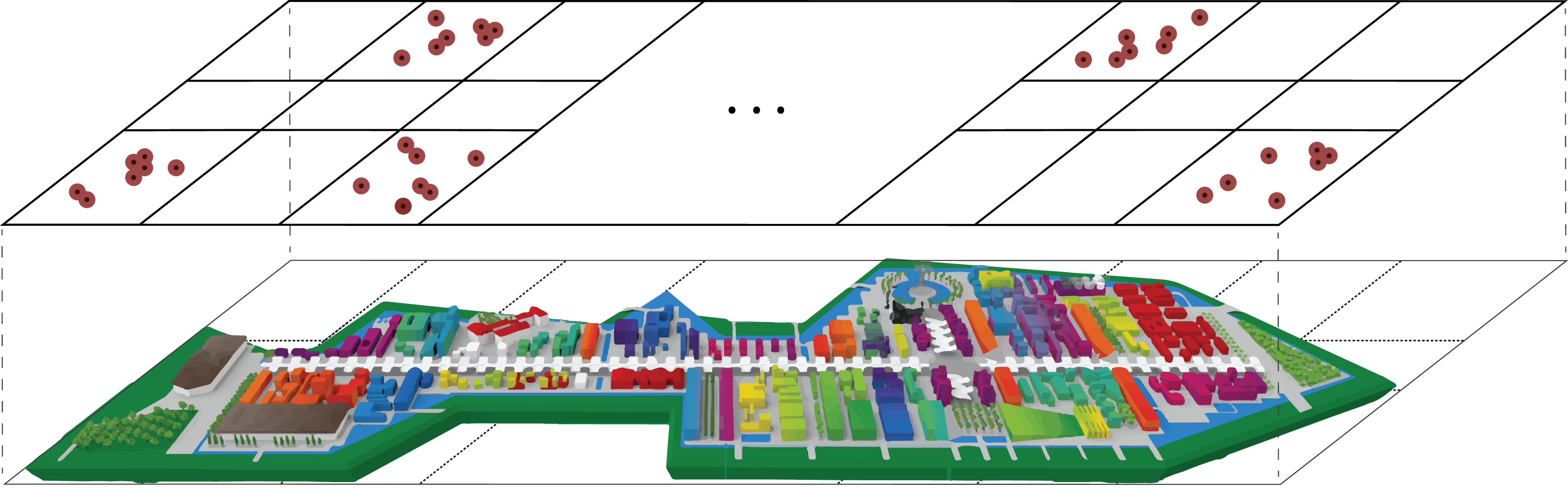}
	\caption{Map of Expo 2015 area with the position of five of the simulated communities.~\cite{arnaboldi2017personalized}}
	\label{fig:expo}
\end{figure}

In order to validate the overall framework, we simulated three realistic scenarios: a big event (Expo 2015), a conference venue (ACM KDD 2015), and a working day in the Helsinki city center.
In this case, we investigated the efficiency of p-PLIERS in a fully decentralized environment, where the contents are dynamically generated over time by nodes. In particular, we measured the similarity of the recommendations generated by the framework on the nodes' LKGs and on the global knowledge graph (GKG), that is, the union of all the LKGs at each time-step of the simulations.
To this aim, we used both real and synthetic mobility traces, and we built three different datasets from Twitter for the nodes' content generation during the simulations.
Specifically, for the Expo 2015 scenario, we simulated different sets of nodes (i.e., 250, 500, and 900 nodes) moving in the Expo area of Milan using the HCMM~\cite{boldrini2010hcmm} human mobility model.
Using this model, we were able to simulate the presence of different communities, which fits well with the reference scenario, as the Expo area was divided into several pavilions.
Figure~\ref{fig:expo} depicts the simulated area in the Expo scenario with the communities of nodes used in HCMM.
In addition, we used the dataset related to the World Food Day in Expo 2015 as content (i.e., tweets) generation pattern of the nodes during the simulations.

As a second dynamic scenario, we considered a school campus during a conference event, where people stay most of the time within rooms, but they regularly gather at breaks (e.g., coffee or lunch breaks). Therefore, we used real contact traces representing the physical interactions of a group of students, professors, and staff of an American high school during a typical school day (i.e., the high resolution Human Contact Network - HCN~\cite{salathe2010high}).
As American high schools are not organized into classes as in Europe, but rather around study tracks, and students are free to decide which lectures to attend, we think that their movements can be assimilated to those of people attending a large conference.
For this reason, we downloaded the tweets generated during a large computer science conference (i.e., the 21st ACM SIGKDD Conference on Knowledge Discovery and Data Mining), and we used this data to simulate the nodes' content generation during the simulation.

Finally, in order to simulate the use of p-PLIERS on a larger scale than the previous scenarios, we extracted the contact traces of a typical working day in Helsinki using a realistic human mobility model highly customized on the considered area.
Specifically, we used the ONE~\cite{keranen2009one} simulator, which implements the \emph{Working Day Mobility Model} (\emph{WDMM})~\cite{ekman2008working} and the map of the city center of Helsinki.
Then, we downloaded the tweets generated within the same geographic area and we used them as content generated by nodes during the simulation.

In the best case, i.e., the conference scenario, the average similarity reaches 80\% already in two hours simulated time, over eight hours duration. In the worst case, i.e., Helsinki scenario, the average similarity reaches 80\% at the end of the simulation, due to the reduced number of contacts during the working day.

This implies that, in dense scenarios, nodes require just few contacts to well approximate the global knowledge with their local graphs and proved that p-PLIERS is able to provide effective recommendations, comparable to those achievable if global knowledge were accessible to nodes.
For the urban scenario of Helsinki, in which the considered area is much larger than that of the other scenarios and the density of nodes is lower, the results indicate that a larger time window is required for a good approximation of the global knowledge about contents in the network. 

With a view to smart cities, a possible solution to improve the diffusion of knowledge and the accuracy of p-PLIERS might be based on exploiting also the public transportation system nodes (e.g., buses, trams, or taxis) as additional information carriers.


In Table~\ref{tab:recsys_msn} we summarise and compare the solutions presented in the previous sections, specifying for each of them the method implemented for recommendation, the context information used to optimize the process, and the details of their performance evaluation in terms of (i) dataset, (ii) mobility traces and (iii) performance metrics. 

\begin{table}[]
	\footnotesize
	\centering
	\caption{RS for MSN}
	\label{tab:recsys_msn}
	\begin{tabular}{|l|l|l|l|l|l|}
	\hline
	\rowcolor[HTML]{EFEFEF} 
	\multicolumn{1}{|c|}{\cellcolor[HTML]{EFEFEF}}                            & \multicolumn{1}{c|}{\cellcolor[HTML]{EFEFEF}}                            & \multicolumn{1}{c|}{\cellcolor[HTML]{EFEFEF}}                              & \multicolumn{3}{c|}{\cellcolor[HTML]{EFEFEF}Evaluation}                                                                                                                                    \\ \cline{4-6} 
	\rowcolor[HTML]{EFEFEF} 
	\multicolumn{1}{|c|}{\multirow{-2}{*}{\cellcolor[HTML]{EFEFEF}Algorithm}} & \multicolumn{1}{c|}{\multirow{-2}{*}{\cellcolor[HTML]{EFEFEF}RS method}} & \multicolumn{1}{c|}{\multirow{-2}{*}{\cellcolor[HTML]{EFEFEF}Context Info}} & \multicolumn{1}{c|}{\cellcolor[HTML]{EFEFEF}Dataset} & \multicolumn{1}{c|}{\cellcolor[HTML]{EFEFEF}Mobility scenario}           & \multicolumn{1}{c|}{\cellcolor[HTML]{EFEFEF}Metrics}                 \\ \hline
	
		OppCF~\cite{de2007collaborative}                                                                       & user-based CF                                                            & \begin{tabular}[c]{@{}l@{}}- location\\ - time\\ - ratings\end{tabular}  & \multicolumn{1}{c|}{-}                            & \multicolumn{1}{c|}{-}                                          & \multicolumn{1}{c|}{-}                                               \\ \hline
	
	Mobhinter~\cite{schifanella2008mobhinter}                                                                 & user-based CF                                                            & - ratings                                                                  & - MovieLens                                       & - Random                                                        & \begin{tabular}[c]{@{}l@{}}- RMSE\\ - MAE\end{tabular}               \\ \hline
	diffeRS~\cite{del2010differs}                                                                   & \begin{tabular}[c]{@{}l@{}}user-based CF\\ + heuristic\end{tabular}      & - ratings                                                                  & - MovieLens                                       & \begin{tabular}[c]{@{}l@{}}- MIT Reality\\ Mining~\cite{mit-reality-20050701}\end{tabular}  & \begin{tabular}[c]{@{}l@{}}- MAE\\ - Coverage\end{tabular}           \\ \hline

	locPref~\cite{zhao2016robust}                                                                   & \begin{tabular}[c]{@{}l@{}}user-based CF\\ + heuristic\end{tabular}      & \begin{tabular}[c]{@{}l@{}}- location\\ - time\\ - ratings\end{tabular}    & - LocShare                                        & - Map-based                                                     & \begin{tabular}[c]{@{}l@{}}- Custom\\ Accuracy\\ metric\end{tabular} \\ \hline \hline
	ICe-Habit~\cite{lo2010folksonomy}                                                                 & tags correlations                                                        & - tags                                                                     & - MovieLens                                       & \begin{tabular}[c]{@{}l@{}}- MIT Reality\\ Mining~\cite{mit-reality-20050701}\end{tabular}  & - Recall                                                             \\ \hline
	p-PLIERS~\cite{arnaboldi2017personalized}                                                                  & \begin{tabular}[c]{@{}l@{}}graph- and\\ diffusion-based\end{tabular}     & \begin{tabular}[c]{@{}l@{}}- users\\ - items\\ - tags\end{tabular}         & - Twitter                                         & \begin{tabular}[c]{@{}l@{}}- HCMM~\cite{boldrini2010hcmm}\\- HCN~\cite{salathe2010high}\\- WDMM~\cite{ekman2008working}\end{tabular} & \begin{tabular}[c]{@{}l@{}}- Precision\\ - Recall\end{tabular}       \\ \hline
	\end{tabular}
	\end{table}

\section{Concluding remarks and open challenges}
\label{sec:conclusions}

In this work, we have presented a survey of the main RS proposed in the literature for Online and Mobile Social Networks, with particular attention to the use of social context information to improve the recommendation process. We described advantages and drawbacks of standard recommendation techniques in these environments and we highlighted the RS challenges in the fully distributed environment of Mobile Social Networks.  
Several solutions have been proposed for OSN, while the study of efficient RS for MSN is still in its infancy and presents several open research challenges, like the use of additional context information, specifically related to the mobile environment, to further optimize the recommendation process.

	Due to their sensing capabilities and the constant presence in human daily life, modern personal devices (e.g., smartphones and tablets) represent the bridge between the cyber and physical worlds, characterizing the situation in which the user is involved during everyday life.
	Therefore, mining the information provided by these devices, and properly combining them with OSN data, can provide a more accurate model of the user's context and preferences.
	For example, the social context of the user may be defined combining the virtual social relationships extracted from OSNs with the physical contacts among the devices and the information contained in the user's personal device (e.g., the contacts saved in the address book, call logs, and messages).
	In this case, the user's preferences and needs can be inferred from heterogeneous sources of data, e.g., the web browser history, the actions performed in OSNs, the generated contents both in OSN and MSN, the visited locations, and the set of most used mobile apps.
	
	The availability of this heterogeneous information represents a crucial aspect in building and evaluating new recommending solutions.
	However, in order to validate and evaluate the possible solutions in realistic environments, it is necessary to generate new datasets, collecting all the contextual information that can characherizes the mobile environment and can be exploited in MSN.
	Collecting new datasets is a very time consuming task, requiring the implementation of a real prototype application and the involvement of a high number of mobile users in real environments.
	A possible solution is represented by the use of synthetic (but realistic) context data generator (e.g.,~\cite{pasinato2013generating, del2017datagencars})
	but, at the moment, none of these solutions is able to integrate realistic temporal settings in the data generation process.
	Since user's preferences are highly dynamic, especially in a mobile environment, the use of an obsolete preference model could negatively affect the quality of the recommendations.

	In addition, several frameworks have been recently proposed to evaluate RS performance in OSN, e.g., \emph{LibRec}~\cite{guo2015librec}, \emph{LensKit}~\cite{ekstrand2011rethinking}, \emph{RankSys}~\cite{castells2015novelty}, and \emph{CARSKit}~\cite{zheng2015carskit}. They provide an implementation of some RS methods (e.g., CF, MF) and performance metrics, but they are developed exclusively for centralized RS, not considering the unique characteristics of MSNs (e.g, the limited knowledge available in each device).
	Therefore, in order to perform a fair comparison among different RS for MSN, we should define a common evaluation framework with the following three main characteristics:

\begin{itemize}
	\item \emph{Mobility}: the RSs should be evaluated in realistic mobile scenarios.
	Therefore, the evaluation framework should include datasets of real mobility traces and human mobility models.

	\item \emph{Data}: as we previously pointed out, RS for MSN should characterize the user's context with several heterogeneous data.
	To this aim, the framework should also include proper datasets or context data generators.
	In addition, given the extremely dynamic nature of the MSN environment, during the simulation nodes should be able to generate data over time.

	\item \emph{Evaluation}: the framework should include the main evaluation metrics proposed in the literature for the two main recommendation tasks (i.e., prediction and ranking).
	Furthermore, due to the typically limited duration of the opportunistic contacts, it is also needed to consider the time complexity of the RS in MSN.
	In fact, if the RS is not able to provide recommendations in a limited time window, a node may miss the opportunity to exchange useful data with another one.
\end{itemize}

All these aspects represent a wide research area, including not only the definition of new RS algorithms, but also the creation of experimental testbeds including new opportunistic networking protocols, opportunistic sensing features and context reasoning tools.



\section{Acknowledgements}
This work was carried out in the framework of the INTESA project, co-funded by the Tuscany Region (Italy) under the Regional Implementation Programme for Underutilized Areas Fund (PAR FAS 2007-2013) and the Research Facilitation Fund (FAR) of the Ministry of Education, University and Research (MIUR).

\section{References}

\bibliography{paper}
	
\end{document}